\documentclass[11pt,onecolumn]{article}
\usepackage{geometry,setspace}
\usepackage[font=small,format=plain,labelfont=bf,justification=raggedright,singlelinecheck=false]{caption}
\usepackage{hyperref,cite}
\usepackage{amsmath,amsfonts,amssymb,amsthm,upgreek,mathrsfs,bm}
\usepackage{graphicx,subfig,tikz}
\usepackage{physics}
\usepackage{extarrows}
\usepackage{orcidlink}

\hypersetup{colorlinks=true,linkcolor=blue,anchorcolor=blue,citecolor=blue,urlcolor=blue}
\allowdisplaybreaks[2]

\numberwithin{equation}{section}

\title{Missing Descendants in the Carrollian Conformal Family}
\author{Yu-fan Zheng\orcidlink{0000-0001-7405-582X}$^{a}$\footnote{\href{mailto:zhengyufan@bimsa.cn}{zhengyufan@bimsa.cn}}}
\date{\today}

\begin{document}

	\maketitle
\begin{center}
    {\it
        $^{a}$ Beijing Institute of Mathematical Sciences and Applications (BIMSA), Huaibei Town, Huairou District, Beijing 101408, China \\
    }
    \vspace{10mm}
\end{center}

	\begin{abstract}
		In the Carrollian conformal algebra, the relation $[P^0,K^0]=0$ implies that $K^0$ annihilates all operators generated from a primary by temporal translation $P^0$.
		The standard conformal family defined by translation descendants does not contain the independent descendant chain that $K^0$ lowers successively to the primary.
		We construct the complete Carrollian conformal representation by including this chain together with all of its translation descendants.
		We derived the corresponding local operators, orbit structures, and checked their Casimirs in 2D and 3D.
		For each orbit configuration, the global two-point Ward identities fix the kinematic factors and selection rules for both magnetic non-contact branches and electric contact branches.
		Unlike ordinary conformal symmetry, Carrollian conformal symmetry generally determines the correlators only up to arbitrary functions of Carrollian invariants rather than constants.
		The complete representation contains sectors with $\mathcal{C}_2=m^2=\kappa\rho-\beta^2>0$ for 2D and $\mathcal{C}_2=m^2=\kappa\rho-\vec{\beta}^{\,2}>0$ for 3D, which may describe massive states in flat holography. 
	\end{abstract}

	\tableofcontents

\section{Introduction}

	Carrollian symmetry was found independently by L\'evy-Leblond and Sen Gupta through the ultra-relativistic contraction of Poincar\'e symmetry \cite{Levy-Leblond:1965,Gupta:1966}, and it was later included among the possible kinematical symmetries of spacetime \cite{Bacry:1968zf}.
	A Carrollian boost leaves the spatial coordinates fixed and shifts the time coordinate by a spatially dependent amount.
	In the same limit, the light cone collapses to the time direction and the metric becomes degenerate along that direction.
	Later work developed the geometric and conformal structures associated with this limit \cite{Duval:2014uoa,deBoer:2023fnj,Ciambelli:2025unn}. \par

	Carrollian field theories provide dynamical realizations of this kinematics.
	They can be obtained from ultra-relativistic limits of relativistic theories or constructed directly from Carrollian diffeomorphism and Weyl invariance \cite{Bagchi:2019xfx,Gupta:2020dtl,Hao:2021urq,Hao:2022xhq}.
	The relation between Carrollian and Bargmann geometry gives another construction through null hypersurfaces \cite{Duval:2014uoa,Duval:2014lpa}.
	Null reduction of Bargmann invariant theories produces off-shell Carrollian actions and relates the electric and magnetic sectors to different parent theories in one higher dimension \cite{Chen:2023pqf}.
	These methods have produced scalar, fermionic, and gauge theories with explicit actions and correlation functions.
	Particle models provide complementary examples.
	A free Carrollian particle does not move in space \cite{Levy-Leblond:1965,Bergshoeff:2014jla}, while interacting conformal Carroll particles can have nontrivial dynamics \cite{Casalbuoni:2023bbh}.
	Their orbit structure and quantization also connect some Carrollian systems to fracton models \cite{Figueroa-OFarrill:2023vbj,Figueroa-OFarrill:2023qty}. \par

	When conformal symmetry is present, local fields and operators form representations of the Carrollian conformal algebra.
	These representations contain primaries and descendants as in an ordinary CFT, while Carrollian boosts can mix finite or infinite multiplets.
	Explicit field theories provide examples of chain representations and staggered modules \cite{Chen:2021xkw,Chen:2023pqf}.
	The representation data enter the Ward identities and determine the invariant dependence and distributional support allowed in correlation functions.
	Known solutions include electric contact correlators and magnetic non-contact correlators \cite{Chen:2021xkw,Bagchi:2023fbj,Salzer:2023jqv,Cotler:2024xhb}. \par

	The main application relevant here is null infinity.
	Its null generators define the degenerate direction of a Carrollian structure, while transverse sections carry the spatial metric \cite{Ciambelli:2018ojf,Donnay:2019jiz,Bagchi:2019clu,Chen:2023pqf}.
	The asymptotic symmetry of flat spacetime at null infinity is the Bondi--Metzner--Sachs symmetry \cite{Bondi:1962px,Sachs:1962wk,Barnich:2010eb,Barnich:2011mi}.
	The BMS algebra is an infinite dimensional extension of Carrollian conformal symmetry \cite{Duval:2014uva,Duval:2014lpa,Afshar:2024llh}.
	Scalar, vector, and gravitational fields at future null infinity have been analyzed through their boundary degrees of freedom and symmetry generators \cite{Liu:2022mne,Liu:2023qtr,Liu:2023gwa}.
	Carrollian CFT therefore provides a natural boundary language for flat holography even before a complete bulk holographic dictionary is known \cite{Bagchi:2025vri,Ruzziconi:2026bix}. \par

	The boundary representations used in flat holography have developed differently in 2D and 3D.
	For a 2D boundary, the relation between BMS$_3$ and the infinite dimensional Carrollian conformal algebra was first formulated through the correspondence between BMS$_3$ and the Galilean conformal algebra (GCA) \cite{Bagchi:2010zz}.
	The centrally extended BMS$_3$ algebra and flat limits of 3D gravity provided further foundations for this description \cite{Barnich:2006av,Barnich:2012aw,Bagchi:2012yk}.
	The resulting 2D theories admit discrete boost multiplets and continuous boost charge representations.
	Their correlators can depend on the boost charge, and their quadratic Casimir can be nonzero \cite{Bagchi:2012cy,Bagchi:2016geg,Bagchi:2017cpu,Chen:2020vvn,Chen:2021xkw,Chen:2022jhx}.
	A nonzero boost charge has also been used to construct Carrollian conformal bases for massive and tachyonic particles in 3D \cite{Liu:2026toc}.
	Induced and coadjoint representations of BMS$_3$ provide complementary descriptions based on group orbits \cite{Barnich:2014kra,Barnich:2015uva}. 
	For a 3D boundary relevant to 4D flat holography, the commonly used highest weight representations describe massless radiation and contain no independent boost charge vector \cite{Bagchi:2016bcd,Chen:2021xkw,Salzer:2023jqv}.
	Their correlation functions contain the familiar electric contact and magnetic non-contact branches \cite{Chen:2021xkw,Bagchi:2023fbj,Salzer:2023jqv,Cotler:2024xhb}. \par

	At null infinity, the boundary representations and their Ward identities organize the field data and constrain correlation functions \cite{Bagchi:2016bcd,Chen:2021xkw,Saha:2023hsl,Salzer:2023jqv}.
	They also relate null infinity correlators to scattering amplitudes and soft theorems \cite{Strominger:2013jfa,He:2014laa,Bagchi:2022emh,Mason:2023mti}, while Carrollian amplitudes can be defined directly in general dimensions \cite{Liu:2024llk}.
	Holographic currents and gravitational balance laws provide another class of observables \cite{Ruzziconi:2024kzo,Fiorucci:2025twa}.
	Recent work has extended the analysis to operator products, Ward identities in momentum space, and higher-point functions in general dimensions \cite{Nguyen:2025sqk,Kulkarni:2025qcx,Marotta:2025qjh}.
	Every such calculation begins with a choice of boundary representation, so its scope determines which operator sectors and observables can appear. \par

	The local Carrollian conformal family usually defined in the literature contains the descendants generated by translations. However, none of these descendants is mapped back to the primary operator by the action of the $K^0$ generator \cite{Bagchi:2016bcd,Chen:2021xkw,Salzer:2023jqv}.
	In an ordinary conformal algebra, temporal translation $P^0$ generates a descendant chain that the temporal special conformal generator $K^0$ lowers toward the primary.
	The Carrollian conformal algebras in eq. \eqref{eq:TwoDimensionalCarrollianAlgebra} and \eqref{eq:ThreeDimensionalCarrollianAlgebra} instead satisfy $[P^0,K^0]=0$.
	If a primary $\mathcal{O}$ is annihilated by $K^0$, this relation gives $K^0(P^0)^n\mathcal{O}=0$.
	The descendants generated by $P^0$ therefore cannot be lowered to $\mathcal{O}$ by $K^0$.
	We call the independent chain that is lowered successively by $K^0$ the missing descendants in the sense that its operators are absent from that family definition.
	A second limitation is that the standard 3D representation used for massless radiation lies in the sector
	\begin{equation}
		\mathcal{C}_2=0.
	\end{equation}
	It therefore does not contain the sectors in which a positive mass can be identified with a nonzero quadratic Casimir \cite{Liu:2026toc}. \par

	We construct the local conformal family that contains both the descendants generated by translations $P^{\mu}$ and the independent $K^0$ descendants.
	We introduce a sequence $\mathcal{O}_n$ satisfying the $K^0$ chain in eq. \eqref{eq:K0DescendantChain}, and include the translation descendants of every $\mathcal{O}_n$ in eq. \eqref{eq:CompleteDescendantFamily}.
	The $K^0$ descendant chains and the boost multiplets are packaged into the generating operators
	\begin{equation}
		\mathcal{O}_{\Delta}(u,z;\beta,\kappa),\qquad \mathcal{O}_{\Delta,l}(u,z^a;\beta^a,\kappa),
	\end{equation}
	for 2D and 3D, respectively.
	The pairs $(u,z)$ and $(u,z^a)$ are Carrollian spacetime coordinates.
	The labels $\Delta$ and $l$ specify scaling and rotation, while $\beta$ and $\beta^a$ organize the boost multiplets and $\kappa$ organizes the $K^0$ descendant chain.
	We call this family complete since it includes the independent $K^0$ descendant chain together with its translation descendants.
	The generator action at the origin, followed by translation to a general spacetime point, defines the local operators and yields the differential representations in eq. \eqref{eq:TwoDimensionalRepresentation} and \eqref{eq:ThreeDimensionalRepresentation}. \par

	The complete representation has the same basic structure in 2D and 3D.
	The symmetry transformations divide the representation space into three types of orbits: one with $\kappa>0$, one with $\kappa<0$, and a family with $\kappa=0$, labeled by fixed $\beta$ in 2D and fixed $\vec{\beta}^{\,2}$ in 3D.
	Using the anti-Hermitian condition $G^\dagger=-G$ for symmetry generators, we find that the representation parameters $\beta, \kappa$ are real in 2D, and $\beta^a, \kappa$ are real in 3D. 
	For $\kappa\neq0$ orbits, the conformal dimensions lie on the principal series $\Delta = \frac{d}{2}+i\mathbb{R}$, whereas for $\kappa=0$, the principal series is shifted to $\Delta = \frac{d-1}{2}+i\mathbb{R}$.
	% For generic $\kappa\neq0$, When $\kappa$ is an independent label, the dual dimensions are $3-\Delta$ in 2D and $4-\Delta$ in 3D, with $\operatorname{Re}\Delta=\frac{3}{2}$ and $\operatorname{Re}\Delta=2$, respectively.
	% On the $\kappa=0$ sectors, the pairing does not include $d\kappa$.
	% The corresponding dual dimensions are $2-\Delta$ in 2D and $3-\Delta$ in 3D, with $\operatorname{Re}\Delta=1$ and $\operatorname{Re}\Delta=\frac{3}{2}$.
	% Hermitian conjugation reverses $\beta$ and $\kappa$ in 2D, and it reverses $\beta^a$, $\kappa$, and $l$ in 3D.
	% Finite transformations preserve the sign of $\kappa$.
	% The regions $\kappa>0$ and $\kappa<0$ each form one orbit.
	% At $\kappa=0$, the orbits are labeled by $\beta$ in 2D and by the rotation invariant $\vec{\beta}^{\,2}$ in 3D. \par
	In Fourier space, with $\rho$ conjugate to $u$, the quadratic Casimir is $\mathcal{C}_2=\rho\kappa-\beta^2$ in 2D and $\mathcal{C}_2=\rho\kappa-\vec{\beta}^{\,2}$ in 3D.
	It first decomposes the representation space into sectors of fixed $\mathcal{C}_2$.
	For $\kappa\neq0$, the higher Casimirs contain derivative terms, and the corresponding orbit representations are reducible.
	At $\kappa=0$, the higher Casimirs are constant on each fixed orbit.
	The stabilizer analysis then gives an irreducible orbit representation when no additional internal labels are present.
	The known 2D representations with boost charge are recovered in this sector with $\xi = i\beta$, while the point $\vec{\beta}=\kappa=0$ gives the standard representation used in 4D flat holography. \par

	A related sequence $|\Delta+n\rangle$ was recently obtained by decomposing a relativistic conformal Verma module into celestial conformal primaries and then taking its magnetic Carrollian limit \cite{Bekaert:2026cib}.
	In that construction, $K^0$ lowers the sequence, while temporal translations and boosts connect adjacent levels through descendants generated by spatial translations.
	These relations define a particular indecomposable Poincar\'e module and leave its quadratic Casimir at $\mathcal{C}_2=0$.
	Here the $\mathcal{O}_n$ sequence is introduced directly as the missing $K^0$ descendant chain of the Carrollian conformal family.
	Its boost and translation descendants are independent, and the resulting representation contains sectors with $\mathcal{C}_2\neq0$.
	The sectors with positive $\mathcal{C}_2$ provide the boundary representation space used in a dictionary for massive particles in flat holography \cite{workinprogress}. \par

	We then solve the global Ward identities for two-point correlation functions.
	The candidate invariants $\mathcal{I}_A$ are invariant under the finite Carrollian transformations in Table \ref{tab:TwoDimensionalCompletedFiniteTransformations} and \ref{tab:ThreeDimensionalCompletedFiniteTransformations}.
	The solutions take the schematic form
	\begin{equation}
		\mathcal{G}=f(\mathcal{I}_A)\,\mathcal{K}\,\mathcal{S}.
	\end{equation}
	The Ward identities determine the kinematic factor $\mathcal{K}$, the distributional support $\mathcal{S}$, and the selection rules.
	When continuous invariants remain, the arbitrary function $f$ contains information that global symmetry does not fix.
	This differs from an ordinary CFT two-point function, which is fixed up to an overall coefficient.
	We classify the correlators according to whether both operators belong to $\kappa\neq0$ orbits, both belong to $\kappa=0$ orbits, or the pair is mixed.
	For two operators in $\kappa\neq0$ orbits, the Ward identities admit a generalized magnetic non-contact branch with shifted spatial support and an independent electric contact branch.
	Both branches can retain arbitrary functions of the invariants, and the contact branch imposes no relation between $\Delta_1$ and $\Delta_2$.
	When both operators belong to $\kappa=0$ orbits, the magnetic branch has a fixed spatial power law, while its remaining functional freedom depends only on the boost orbit.
	A mixed pair admits only a magnetic non-contact solution.
	Its contact ansatz reduces to the sector with both $\kappa_i=0$. \par

	The 3D contact solutions also illustrate the importance of the underlying space.
	Point support on the ordinary coordinate and representation planes retains no angular information, so the rotation Ward identity requires $l_1+l_2=0$.
	A blow-up of the origin to $S^1$ can replace point support in a 2D plane by a circle that records the direction of approach.
	The angular factors can then compensate $l_1+l_2$.
	For the $\kappa=0$ contact branch, two angle differences on the enlarged space are also Carrollian invariants and can enter an arbitrary function.
	These solutions are distributions on enlarged spaces and are not alternative expressions for distributions on the original planes. \par

	The two-point function of a local operator and its conjugate determines the pairing of the corresponding states.
	For $\kappa\neq0$, the conjugate-pair correlators contain Dirac distributions in the spatial separation and the continuous representation labels.
	The local operators on these orbits are therefore delta-function normalized.
	For $\kappa=0$, a conjugate pair with nonzero boost labels has a vanishing two-point function, so the corresponding state has zero norm.
	The nonzero conjugate pairings in this sector are supported at $\beta=0$ in 2D and $\vec{\beta}=0$ in 3D, and hence select the zero modes of both the boost and $K^0$ descendant expansions.
	These zero modes admit an electric contact pairing and an additional magnetic non-contact pairing. \par

	The remainder of the paper is organized as follows.
	Section \ref{sec:CompleteCarrollianRepresentations} constructs the complete local representations in 2D and 3D and determines their properties, including reality conditions, duals, finite transformations, orbit decompositions, and Casimir structures.
	Section \ref{sec:CorrelationFunctions} solves the Ward identities for two-point correlation functions in the three $\kappa$ configurations and identifies the magnetic non-contact and electric contact branches.
	Section \ref{sec:Discussion} summarizes the results and discusses future directions. \par

\section{Complete Carrollian Conformal Representation}\label{sec:CompleteCarrollianRepresentations}

	This section constructs the complete Carrollian conformal family from the descendant chains.
	The relation $[P^0,K^0]=0$ separates descendants generated by temporal translations from descendants lowered successively by $K^0$.
	Subsection \ref{subsec:MissingDescendants} first explains the mechanism common to the 2D and 3D theories and introduces an independent $K^0$ chain together with its translation descendants. 
	We then realize the same construction independently in 2D and 3D.
	In each dimension, the discrete boost chain is resummed into a continuous boost charge, while the discrete $K^0$ chain is resummed into the representation parameter $\kappa$.
	After fixing the generator action at the origin, the Baker--Campbell--Hausdorff formula determines the local operator at a general Carrollian point.
	Subsections \ref{subsec:TwoDimensionalRepresentation} and \ref{subsec:ThreeDimensionalRepresentation} then determine the properties, including reality conditions, dual representations, finite transformations, orbit decompositions, and Casimir structures.
	% The complete representation reduces to the boost charge representation at $\kappa=0$.
	% In 3D, the point $\vec{\beta}=\kappa=0$ gives the standard representation used in 4D flat holography. \par

\subsection{The Missing \texorpdfstring{$K^0$}{K0} Descendants}\label{subsec:MissingDescendants}

	A conformal family forms a highest weight representation of the conformal algebra.
	Its primary operator $\mathcal{O}_{\mathrm{p}}$ satisfies
	\begin{equation}
		[D,\mathcal{O}_{\mathrm{p}}]=\Delta\mathcal{O}_{\mathrm{p}}, \qquad [K^\mu,\mathcal{O}_{\mathrm{p}}]=0.
	\end{equation}
	Focusing on the temporal direction, its descendants are generated by repeated actions of $P^0$ and take the form%
		\footnote{For $m\geq2$, $G_1\cdots G_m\mathcal{O}$ denotes the nested commutator $[G_1,[G_2,\ldots,[G_m,\mathcal{O}]\ldots]]$.}
	\begin{equation}
		(P^0)^n\mathcal{O}_{\mathrm{p}}=\partial_0^n\mathcal{O}_{\mathrm{p}}.
	\end{equation}
	In an ordinary conformal algebra, $[K^0,P^0]$ is proportional to $D$, and therefore for the descendants,
	\begin{equation}
		K^0(P^0)^n\mathcal{O}_{\mathrm{p}}\propto(P^0)^{n-1}\mathcal{O}_{\mathrm{p}}.
	\end{equation}
	Thus, in an ordinary conformal family, generation by $P^\mu$ and successive reduction under $K^\mu$ provide equivalent characterizations of the descendant structure. \par
	
	By contrast, the family definition usually used for the Carrollian conformal algebra includes the descendants generated by translations, while the temporal direction permits a second descendant structure.
	The 2D and 3D Carrollian conformal algebras given in \eqref{eq:TwoDimensionalCarrollianAlgebra} and \eqref{eq:ThreeDimensionalCarrollianAlgebra} share the relation
	\begin{equation}
		[P^0,K^0]=0.
	\end{equation}
	Consequently,
	\begin{equation}
		K^0(P^0)^n\mathcal{O}_{\mathrm{p}}=(P^0)^nK^0\mathcal{O}_{\mathrm{p}}=0.
	\end{equation}
	The descendants $(P^0)^n\mathcal{O}_{\mathrm{p}}$ generated by temporal translations are therefore annihilated by $K^0$ and cannot be successively lowered back to the primary.
	The two equivalent characterizations of descendants in an ordinary conformal family thus separate in the temporal direction of the Carrollian conformal algebra. \par

\begin{figure}
		\centering
		\captionsetup{width=0.8\linewidth}
		\includegraphics[width=0.75\linewidth]{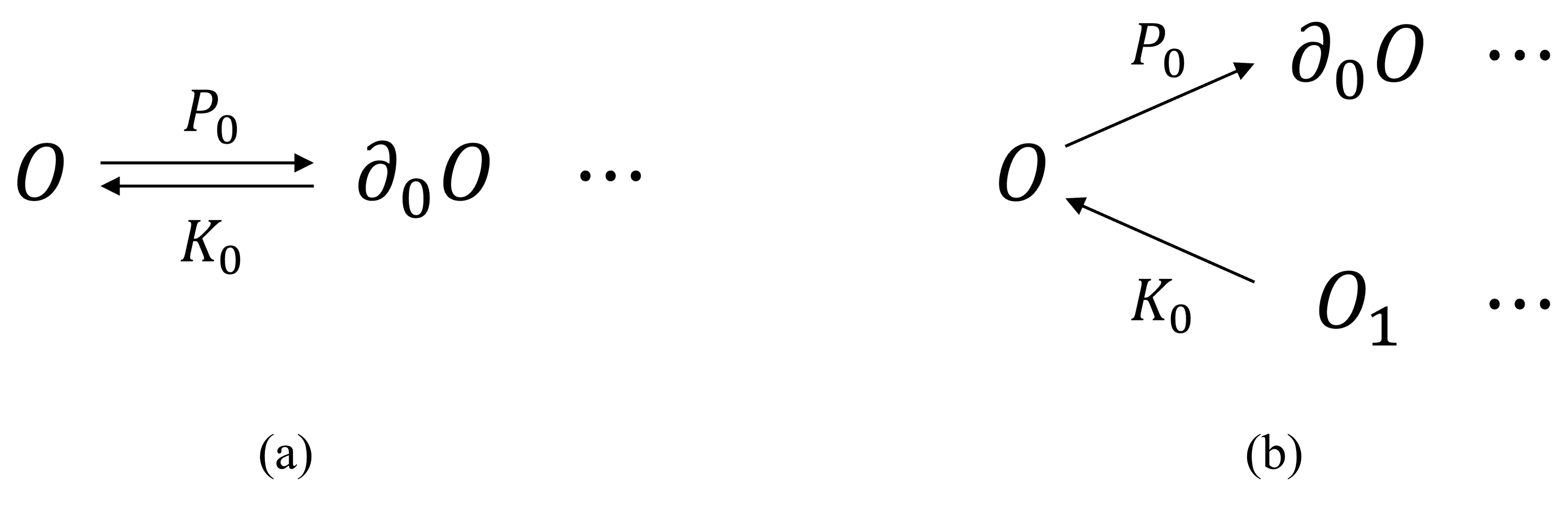}
		\caption{Descendant structures in (a) an ordinary conformal family and (b) a Carrollian conformal family.
			In (a), $P^0$ and $K^0$ act in opposite directions along the same chain.
			In (b), the translation chain and the independent chain lowered by $K^0$ define separate descendant directions.}
		\label{fig:K0Descendants}
	\end{figure}

	We include the descendants that return to the primary under the action of $K^0$ by introducing a sequence of operators $\mathcal{O}_n$, normalized such that
	\begin{equation}\label{eq:K0DescendantChain}
		\mathcal{O}_0=\mathcal{O}_{\mathrm{p}},\qquad [K^0,\mathcal{O}_n]=\mathcal{O}_{n-1},\qquad n\geq1.
	\end{equation}
	The commutator $[D,K^0]=-K^0$ then fixes their scaling dimensions to be
	\begin{equation}
		[D,\mathcal{O}_n]=(\Delta+n)\mathcal{O}_n.
	\end{equation}
	Their dimensions are $\Delta+n$, and $K^0$ lowers them successively to the primary.
	They therefore form the independent descendant chain missing from the family generated by translations. 
	Figure \ref{fig:K0Descendants} makes manifest that the difference lies not in the existence of temporal translation descendants, but in whether $K^0$ acts as a lowering operator on the same descendant chain.\par

	% Each $\mathcal{O}_n$ denotes the complete boost multiplet at that level.
	We include the translation descendants
	\begin{equation}\label{eq:CompleteDescendantFamily}
		P^{\mu_1}\cdots P^{\mu_m}\mathcal{O}_n =\partial_{\mu_1}\cdots\partial_{\mu_m}\mathcal{O}_n, \qquad m,n\geq0.
	\end{equation}
	The usual descendants of the primary constitute the $n=0$ part of this family, while the sectors with $n\geq1$ contain the additional $K^0$ descendant chains together with all of their translation descendants.
	% Our complete Carrollian conformal representation consists of the independent $\mathcal{O}_n$ tower and all translation descendants displayed in eq. \eqref{eq:CompleteDescendantFamily}.
	This separation of the two descendant structures follows only from the special temporal commutator and is therefore independent of the number of spatial dimensions.
	In the rest of this section, we realize this mechanism independently in 2D and 3D. \par

\subsection{2D Complete Carrollian Representation}\label{subsec:TwoDimensionalRepresentation}

\subsubsection{Construction of the complete local representation}

	The 2D Carrollian conformal algebra is generated by $\{P^0,P^1,D,B^1,K^0,K^1\}$.
	With $\mu=0,1$, its independent commutators are
	\begin{equation}\label{eq:TwoDimensionalCarrollianAlgebra}
		\begin{aligned}
			&[D,P^\mu]=P^\mu,\qquad [D,K^\mu]=-K^\mu,\qquad [D,B^1]=0,\\
			&[B^1,P^1]=P^0,\qquad [B^1,K^1]=K^0,\qquad [B^1,P^0]=[B^1,K^0]=0,\\
			&[P^0,K^1]=-2B^1,\qquad [P^1,K^0]=2B^1,\qquad [P^1,K^1]=-2D, \\
			&[P^0,P^1]=[P^0,K^0]=[K^0,K^1]=0,
		\end{aligned}
	\end{equation}
	As explained in the previous subsection, the relation $[P^0,K^0]=0$ separates the descendants generated by temporal translation $P^0$ from the $K^0$ descendant direction included in our construction. \par

	We first organize the boost multiplet by considering a sequence of operators $\mathcal{O}_{\Delta}(n_B)$.
	We normalize this boost chain by
	\begin{equation}
		[B^1,\mathcal{O}_{\Delta}(n_B)]=\mathcal{O}_{\Delta}(n_B-1), \qquad [B^1,\mathcal{O}_{\Delta}(0)]=0.
	\end{equation}
	This discrete chain can be organized into the generating operator
	\begin{equation}\label{eq:2DBoostResummation}
		\mathcal{O}_{\Delta}(\beta) = \sum_{n_B=0}^{\infty}(i\beta)^{n_B}\mathcal{O}_{\Delta}(n_B),
	\end{equation}
	which satisfies
	\begin{equation}
		[B^1,\mathcal{O}_{\Delta}(\beta)] = i\beta\mathcal{O}_{\Delta}(\beta).
	\end{equation}
	The factor of $i$ is introduced for later convenience when imposing the reality conditions. \par

	We next introduce the $K^0$ descendant chain following the construction in subsection \ref{subsec:MissingDescendants}.
	Since $[B^1,K^0]=0$, the action of $K^0$ does not change the index along the boost chain.
	We therefore define the basis $\mathcal{O}_{\Delta}(n_B,n_{K^0})$ with two indices by
	\begin{equation}
		\begin{aligned}
			& \mathcal{O}_{\Delta}(n_B,0)=\mathcal{O}_{\Delta}(n_B), \\
			& [B^1,\mathcal{O}_{\Delta}(n_B,n_{K^0})] =\mathcal{O}_{\Delta}(n_B-1,n_{K^0}),\\
			& [K^0,\mathcal{O}_{\Delta}(n_B,n_{K^0})] =\mathcal{O}_{\Delta}(n_B,n_{K^0}-1).
		\end{aligned}
	\end{equation}
	Any operator carrying a negative index is understood to vanish.
	The two descendant chains are simultaneously resummed into the generating operator
	\begin{equation}
		\mathcal{O}_{\Delta}(\beta,\kappa) = \sum_{n_B,n_{K^0}=0}^{\infty} (i\beta)^{n_B}(i\kappa)^{n_{K^0}} \mathcal{O}_{\Delta}(n_B,n_{K^0}),
	\end{equation}
	Every discrete descendant is recovered from the Taylor coefficients of this generating operator
	\begin{equation}
		\mathcal{O}_{\Delta}(n_B,n_{K^0})
		=\frac{i^{-n_B-n_{K^0}}}
		{n_B!~n_{K^0}!}\left.
			\partial_{\beta}^{n_B}\partial_{\kappa}^{n_{K^0}}\mathcal{O}_{\Delta}(\beta,\kappa)
		\right|_{\beta=\kappa=0},
	\end{equation}
	which satisfies
	\begin{equation}
		[B^1,\mathcal{O}_{\Delta}(\beta,\kappa)] =i\beta\mathcal{O}_{\Delta}(\beta,\kappa),\qquad [K^0,\mathcal{O}_{\Delta}(\beta,\kappa)] =i\kappa\mathcal{O}_{\Delta}(\beta,\kappa).
	\end{equation}
	\par

	The commutator $[B^1,K^1]=K^0$ requires the action of $K^1$ to raise $n_B$ and lower $n_{K^0}$ simultaneously.
	Together with $[K^0,K^1]=0$, the algebra fixes the action to be
	\begin{equation}
		[K^1,\mathcal{O}_{\Delta}(n_B,n_{K^0})] = (n_B+c)\mathcal{O}_{\Delta}(n_B+1,n_{K^0}-1).
	\end{equation}
	Acting with $[B^1,K^1]=K^0$ on $\mathcal{O}_{\Delta}(0,n_{K^0})$ fixes $c=1$.
	Therefore,
	\begin{equation}
		[K^1,\mathcal{O}_{\Delta}(n_B,n_{K^0})] = (n_B+1)\mathcal{O}_{\Delta}(n_B+1,n_{K^0}-1),
	\end{equation}
	which becomes
	\begin{equation}
		[K^1,\mathcal{O}_{\Delta}(\beta,\kappa)] = \kappa\partial_{\beta}\mathcal{O}_{\Delta}(\beta,\kappa)
	\end{equation}
	in the continuous basis. \par

	The boost action does not change the scaling dimension because $[D,B^1]=0$, whereas each $K^0$ descendant raises it by one.
	Consequently,
	\begin{equation}
		[D,\mathcal{O}_{\Delta}(n_B,n_{K^0})] = (\Delta+n_{K^0})\mathcal{O}_{\Delta}(n_B,n_{K^0}),
	\end{equation}
	or equivalently,
	\begin{equation}
		[D,\mathcal{O}_{\Delta}(\beta,\kappa)] = \left(\Delta+\kappa\partial_{\kappa}\right) \mathcal{O}_{\Delta}(\beta,\kappa).
	\end{equation}
	The states with $n_{K^0}=0$ form the Carrollian primary boost multiplet, whose lowest component in the boost chain is $\mathcal{O}_{\Delta}(0,0)$.
	By contrast, $\mathcal{O}_{\Delta}(\beta,\kappa)$ organizes the complete family and, for nonzero $\kappa$, is not annihilated by either $K^0$ or $K^1$.
	% Here the spacetime point remains fixed at $(u,z)=(0,0)$, while $\beta$ and $\kappa$ are continuous labels of the representation space. \par

	A local operator at a general spacetime point is obtained by translation.
	Defining
	\begin{equation}
		U(u,z)=\exp\left(uP^0+zP^1\right),
	\end{equation}
	we set
	\begin{equation}
		\mathcal{O}_{\Delta}(u,z;\beta,\kappa) = U(u,z)\mathcal{O}_{\Delta}(\beta,\kappa)U^{-1}(u,z).
	\end{equation}
	The Baker--Campbell--Hausdorff formula then gives
	\begin{equation}\label{eq:TwoDimensionalRepresentation}
		\begin{aligned}
			[P^0,\mathcal{O}_{\Delta}(u,z;\beta,\kappa)] &=\partial_u\mathcal{O}_{\Delta}(u,z;\beta,\kappa), \\
			[P^1,\mathcal{O}_{\Delta}(u,z;\beta,\kappa)] &=\partial_z\mathcal{O}_{\Delta}(u,z;\beta,\kappa), \\
			[D,\mathcal{O}_{\Delta}(u,z;\beta,\kappa)] &=\left(u\partial_u+z\partial_z+\kappa\partial_{\kappa}+\Delta\right) \mathcal{O}_{\Delta}(u,z;\beta,\kappa), \\
			[B^1,\mathcal{O}_{\Delta}(u,z;\beta,\kappa)] &=\left(z\partial_u+i\beta\right) \mathcal{O}_{\Delta}(u,z;\beta,\kappa), \\
			[K^0,\mathcal{O}_{\Delta}(u,z;\beta,\kappa)] &=\left(-z^2\partial_u-2iz\beta+i\kappa\right) \mathcal{O}_{\Delta}(u,z;\beta,\kappa), \\
			[K^1,\mathcal{O}_{\Delta}(u,z;\beta,\kappa)] &=\left(2uz\partial_u+z^2\partial_z +2z\left(\kappa\partial_{\kappa}+\Delta\right) +\kappa\partial_{\beta}+2ui\beta\right) \mathcal{O}_{\Delta}(u,z;\beta,\kappa).
		\end{aligned}
	\end{equation}
	These differential operators realize the complete 2D Carrollian conformal representation on the enlarged space $(u,z;\beta,\kappa)$. \par

	The hypersurface $\kappa=0$ is an invariant sector on which the additional $K^0$ descendant direction is absent.
	On this sector, the generating operator reduces to $\mathcal{O}_{\Delta}(\beta,0)=\mathcal{O}_{\Delta}(\beta)$ and obeys
	\begin{equation}
		[B^1,\mathcal{O}_{\Delta}(\beta,0)]=i\beta\mathcal{O}_{\Delta}(\beta,0).
	\end{equation}
	The boost charge representations in the literature use $[B^1,\mathcal{O}_{\xi}]=\xi\mathcal{O}_{\xi}$ \cite{Bagchi:2012cy,Chen:2020vvn}, so the two conventions are related by $\xi=i\beta$.
	In particular, the continuous boost charge basis is related to the discrete representation introduced in \cite{Chen:2022jhx} through the resummation in \eqref{eq:2DBoostResummation}.
	Thus the $\kappa=0$ sector exactly recovers the boost charge representation, while $\kappa\neq0$ supplies the independent $K^0$ descendant direction. \par

\subsubsection{Properties of the complete representation}\label{subsec:TwoDimensionalCompletedRepresentationProperties}

	In this subsection, we study the reality conditions, dual representation, orbits, and Casimir structure of the complete 2D Carrollian conformal representation. \par

	We use the anti-Hermitian convention $G^\dagger=-G$ for all symmetry generators.
	This condition is inherited from a unitary bulk Poincar\'e representation after identifying the Poincar\'e algebra with the global Carrollian conformal algebra.
	It should be distinguished from the Belavin--Polyakov--Zamolodchikov (BPZ) conjugation defined within a Carrollian conformal field theory (CarrCFT) module \cite{Zheng:2025rfe, Liu:2026toc}.
	% Explicit Carrollian field theories realize their Hermitian structure through canonical quantization \cite{Chen:2024voz,Zheng:2025rfe}.
	For general $\kappa$, we define the formal adjoint using the pairing
	\begin{equation}
		\langle f,g\rangle = \int du\,dz\,d\beta\,d\kappa\, f^*(u,z;\beta,\kappa) g(u,z;\beta,\kappa),
	\end{equation}
	where boundary terms are assumed to vanish.
	Since the differential realization \eqref{eq:TwoDimensionalRepresentation} of $D$ is $u\partial_u+z\partial_z+\kappa\partial_\kappa+\Delta$, imposing $D^\dagger=-D$ gives
	\begin{equation}
		\Delta^\dagger=3-\Delta.
	\end{equation}
	The constant $3$ follows from the identity obtained by integration by parts
	\begin{equation}
		\left\langle f,\left(u\partial_u+z\partial_z+\kappa\partial_\kappa\right)g\right\rangle = -\left\langle \left(u\partial_u+z\partial_z+\kappa\partial_\kappa+3\right)f,g \right\rangle.
	\end{equation}
	It is the divergence $\partial_u u+\partial_z z+\partial_\kappa\kappa=3$, with one contribution from each of the $u$, $z$, and $\kappa$ directions.
	Therefore,
	\begin{equation}
		\Delta=\frac{3}{2}+i\nu, \qquad \nu\in\mathbb{R}.
	\end{equation}
	The conformal dimension lies on the principal line associated with a pairing that includes $d\kappa$.%
		\footnote{In ordinary 2D conformal harmonic analysis and celestial CFT, the principal series has $\operatorname{Re}\Delta=1$ \cite{Pasterski:2017kqt,Karateev:2018oml,Donnay:2020guq}.
			Here the additional scaling direction $\kappa\partial_\kappa$ shifts the anti-Hermitian line to $\operatorname{Re}\Delta=\frac{3}{2}$.}
	Applying the same anti-Hermitian convention to the multiplicative terms in the $B^1$ and $K^0$ actions in \eqref{eq:TwoDimensionalRepresentation} gives
	\begin{equation}
		\beta^\dagger=\beta, \qquad \kappa^\dagger=\kappa.
	\end{equation}
	Thus both $\beta$ and $\kappa$ are real. \par

	Hermitian conjugation maps the representation to its dual.
	For any anti-Hermitian generator $G$,
	\begin{equation}
		\left([G,\mathcal{O}]\right)^\dagger=\left(G\mathcal{O}-\mathcal{O}G\right)^\dagger=\mathcal{O}^\dagger G^\dagger-G^\dagger\mathcal{O}^\dagger=[G,\mathcal{O}^\dagger].
	\end{equation}
	Applying this identity to the defining actions at the origin gives
	\begin{equation}
		\begin{aligned}
			[B^1,\mathcal{O}_\Delta^\dagger(\beta,\kappa)] &=\left(i\beta\mathcal{O}_\Delta(\beta,\kappa)\right)^\dagger =-i\beta\mathcal{O}_\Delta^\dagger(\beta,\kappa), \\
			[K^0,\mathcal{O}_\Delta^\dagger(\beta,\kappa)] &=\left(i\kappa\mathcal{O}_\Delta(\beta,\kappa)\right)^\dagger =-i\kappa\mathcal{O}_\Delta^\dagger(\beta,\kappa).
		\end{aligned}
	\end{equation}
	Thus the dual representation carries the labels $-\beta$ and $-\kappa$.
	Denoting its operator by $\mathcal{O}_{3-\Delta}^\prime$, we obtain
	\begin{equation}
		\mathcal{O}_\Delta^\dagger(u,z;\beta,\kappa) = \mathcal{O}_{3-\Delta}^\prime(u,z;-\beta,-\kappa).
	\end{equation}
	\par

	The invariant $\kappa=0$ sector instead has the pairing
	\begin{equation}
		\langle f,g\rangle_{\kappa=0}=\int du\,dz\,d\beta\,f^*(u,z;\beta,0)g(u,z;\beta,0).
	\end{equation}
	Since $D=u\partial_u+z\partial_z+\Delta$ on this sector, its dual dimension and allowed line are
	\begin{equation}
		\Delta=1+i\nu, \qquad \nu\in\mathbb{R}.
	\end{equation}
	The conjugated operator in this sector is therefore
	\begin{equation}\label{eq:TwoDimensionalZeroKappaDualOperator}
		\mathcal{O}_{\Delta}^{\dagger}(u,z;\beta,0)=\mathcal{O}_{2-\Delta}^{\prime}(u,z;-\beta,0).
	\end{equation}
	\par

	The dual operator determines the bra state paired with the ket created by the original operator acting on the vacuum.
	For $\ket{\mathcal{O}}=\mathcal{O}\ket{0}$, the vacuum inner product with another state created by an operator is
		\begin{equation}
			\langle\mathcal{O}_2\vert\mathcal{O}_1\rangle=\bra{0}\mathcal{O}_2^\dagger\mathcal{O}_1\ket{0}.
		\end{equation}
	The two-point functions derived in Subsection \ref{subsec:TwoDimensionalCorrelators} determine whether this pairing is nonzero and how it is normalized.
	For $\kappa\neq0$, duality pairs the $\kappa>0$ and $\kappa<0$ orbits, and the nonvanishing two-point function between an operator and its dual is supported on the contact branch with coincident spatial support enforced by the Dirac distribution $\delta(z_1-z_2)$.
	The accompanying Dirac distributions in the spatial coordinate and continuous labels show that these states form a generalized basis normalized by delta functions.
	In the $\kappa=0$ sector with $\beta\neq0$, the two-point function between an operator and its dual vanishes, so the corresponding state is null and does not define a normalizable state in this pairing.
	The orbit $\beta=\kappa=0$ admits both an electric contact pairing and an additional magnetic non-contact pairing. \par

	\begin{table}[htbp]
		\renewcommand{\arraystretch}{2}
		\centering
		\caption{Finite transformations of the representation variables in 2D.
			The parameters $u_0$, $z_0$, $\lambda>0$, $b$, $k^0$, and $k^1$ label the corresponding transformations.
			The Fourier variable $\rho$, conjugate to $u$, is introduced in \eqref{eq:TwoDimensionalFourierTransform}.}
		\label{tab:TwoDimensionalCompletedFiniteTransformations}
		\resizebox{\linewidth}{!}{
			\begin{tabular}{c|c|c|c|c|c}
				\hline
				Generator & $u'$ & $z'$ & $\beta'$ & $\kappa'$ & $\rho'$ \\\hline\hline
				$P^0$ & $u+u_0$ & $z$ & $\beta$ & $\kappa$ & $\rho$ \\\hline
				$P^1$ & $u$ & $z+z_0$ & $\beta$ & $\kappa$ & $\rho$ \\\hline
				$D$ & $\lambda u$ & $\lambda z$ & $\beta$ & $\lambda\kappa$ & $\lambda^{-1}\rho$ \\\hline
				$B^1$ & $u+bz$ & $z$ & $\beta$ & $\kappa$ & $\rho$ \\\hline
				$K^0$ & $u-k^0z^2$ & $z$ & $\beta$ & $\kappa$ & $\rho$ \\\hline
				$K^1$
					& $\dfrac{u}{(1-k^1z)^2}$
					& $\dfrac{z}{1-k^1z}$
					& $\beta+\dfrac{k^1\kappa}{1-k^1z}$
					& $\dfrac{\kappa}{(1-k^1z)^2}$
					& $(1-k^1z)^2\rho+2k^1(1-k^1z)\beta+(k^1)^2\kappa$ \\\hline
			\end{tabular}
		}
	\end{table}

	Exponentiating the vector field components of \eqref{eq:TwoDimensionalRepresentation} gives the finite transformations shown in Table \ref{tab:TwoDimensionalCompletedFiniteTransformations}.
	These finite transformations determine the representation orbits.
	For $\kappa\neq0$, translations first set $u=z=0$, a $K^1$ transformation with $k^1=-\beta/\kappa$ then sets $\beta=0$, and a dilatation with $\lambda=1/|\kappa|$ fixes $\kappa$ to its sign.
	The two orbits with nonzero $\kappa$ therefore have the representatives
	\begin{equation}
		(u,z;\beta,\kappa)=(0,0;0,1), \qquad \mathrm{and} \qquad (u,z;\beta,\kappa)=(0,0;0,-1).
	\end{equation}
	When $\kappa=0$, every finite transformation leaves $\beta$ fixed, while translations set $u=z=0$.
	Each fixed value $\beta=\beta_0$ therefore defines an orbit represented by $(0,0;\beta_0,0)$.
	The representation decomposes into
	\begin{equation}
		\kappa>0, \qquad \kappa<0, \qquad \kappa=0\ \mathrm{with}\ \beta=\beta_0.
	\end{equation}
	The duality map $(\beta,\kappa)\mapsto(-\beta,-\kappa)$ exchanges the orbits with $\kappa>0$ and $\kappa<0$ and maps the orbit $\beta=\beta_0$, $\kappa=0$ to the orbit $\beta=-\beta_0$, $\kappa=0$. \par

	For the subsequent Casimir analysis, we Fourier transform the operator in the $u$ direction according to
	\begin{equation}\label{eq:TwoDimensionalFourierTransform}
		\mathcal{O}_\Delta(u,z;\beta,\kappa) =\int\frac{d\rho}{\sqrt{2\pi}}\, e^{-i\rho u} \widetilde{\mathcal{O}}_\Delta(\rho,z;\beta,\kappa),
	\end{equation}
	where $\rho\in\mathbb{R}$.
	It follows that
	\begin{equation}
		\partial_u\longmapsto-i\rho, \qquad u\longmapsto-i\partial_\rho, \qquad u\partial_u\longmapsto-\left(\rho\partial_\rho+1\right).
	\end{equation}
	The Fourier transform preserves the reality conditions $\Delta=\frac{3}{2}+i\nu$ and $\rho,\beta,\kappa,\nu\in\mathbb{R}$.
	Hermitian conjugation reverses $\rho$, so the momentum space dual operator satisfies
	\begin{equation}
		\widetilde{\mathcal{O}}_\Delta^\dagger(\rho,z;\beta,\kappa) =\widetilde{\mathcal{O}}_{3-\Delta}^\prime(-\rho,z;-\beta,-\kappa).
	\end{equation}
	Applying \eqref{eq:TwoDimensionalFourierTransform}, the nontrivial momentum space actions of $D$ and $K^1$ become
	\begin{equation}
		\begin{aligned}
			[D,\widetilde{\mathcal{O}}_\Delta] &=\left(-\rho\partial_\rho+z\partial_z+\kappa\partial_\kappa+\Delta-1\right) \widetilde{\mathcal{O}}_\Delta, \\
			[K^1,\widetilde{\mathcal{O}}_\Delta] &=\left(z^2\partial_z+2z\kappa\partial_\kappa+\kappa\partial_\beta +2\left(\beta-z\rho\right)\partial_\rho+2z(\Delta-1)\right) \widetilde{\mathcal{O}}_\Delta.
		\end{aligned}
	\end{equation}
	The corresponding transformations of $\rho$ are also listed in Table \ref{tab:TwoDimensionalCompletedFiniteTransformations}. \par

	% The finite $K^1$ transformation follows from the flow equations
	% \begin{equation}
	% 	\frac{dz}{dt}=z^2,
	% 	\qquad
	% 	\frac{d\kappa}{dt}=2z\kappa,
	% 	\qquad
	% 	\frac{d\beta}{dt}=\kappa,
	% 	\qquad
	% 	\frac{d\rho}{dt}=2\left(\beta-z\rho\right).
	% \end{equation}
	% Solving these equations with the unprimed variables as the initial values gives
	% \begin{equation}\label{eq:TwoDimensionalFiniteKOne}
	% 	\begin{aligned}
	% 		z'&=\frac{z}{1-tz},
	% 		& \kappa'&=\frac{\kappa}{(1-tz)^2}, \\
	% 		\beta'&=\beta+\frac{t\kappa}{1-tz},
	% 		& \rho'&=(1-tz)^2\rho+2t(1-tz)\beta+t^2\kappa.
	% 	\end{aligned}
	% \end{equation} \par

	The quadratic Casimir of the 2D Carrollian conformal algebra is
	\begin{equation}\label{eq:2DQuadraticCasimir}
		\mathcal{C}_2 =(B^1)^2+P^0K^0 =i\kappa\partial_u-\beta^2 \xrightarrow{\ \mathcal{F}_u\ } \rho\kappa-\beta^2.
	\end{equation}
	Its invariance under $P^0$, $P^1$, $B^1$, and $K^0$ is immediate because these transformations leave $\rho\kappa-\beta^2$ unchanged.
	Under a finite dilatation,
	\begin{equation}
		(\lambda^{-1}\rho)(\lambda\kappa)-\beta^2 = \rho\kappa-\beta^2.
	\end{equation}
	For a finite $K^1$ transformation, direct substitution gives
	\begin{equation}
		\begin{aligned}
			\rho'\kappa'-(\beta')^2 &=\left((1-k^1z)^2\rho+2k^1(1-k^1z)\beta+(k^1)^2\kappa\right)\frac{\kappa}{(1-k^1z)^2}-\left(\beta+\frac{k^1\kappa}{1-k^1z}\right)^2 \\
			&=\rho\kappa-\beta^2.
		\end{aligned}
	\end{equation}
	Thus $\mathcal{C}_2$ is invariant under all finite Carrollian conformal transformations.
	In the momentum space realization, it acts multiplicatively, and the representation space decomposes into eigenspaces of $\mathcal{C}_2$.
	On the $\kappa=0$ sector, $\mathcal{C}_2=-\beta^2=\xi^2$ after using $\xi=i\beta$, so the boost charge representations of \cite{Bagchi:2012cy,Chen:2020vvn,Chen:2022jhx,Chen:2021xkw} can carry a nonzero quadratic Casimir. \par

	The 2D Carrollian conformal algebra also admits the Pauli--Lubanski Casimir
	\begin{equation}
		\mathcal{C}_{\mathrm{PL}} = P^0K^1-K^0P^1-2B^1D.
	\end{equation}
	Substituting the differential representation gives
	\begin{equation}
		\mathcal{C}_{\mathrm{PL}} = \kappa\partial_\beta\partial_u-i\kappa\partial_z -2i\beta\kappa\partial_\kappa+2i\beta(1-\Delta) \xrightarrow{\ \mathcal{F}_u\ } -i\left(\kappa\rho\partial_\beta+\kappa\partial_z +2\beta\kappa\partial_\kappa+2\beta(\Delta-1)\right).
	\end{equation}
	For $\kappa\neq0$, the central operator $\mathcal{C}_{\mathrm{PL}}$ contains derivative terms and the representation on each such orbit is reducible.
	On each $\kappa=0$ orbit with fixed $\beta=\beta_0$, $\mathcal{C}_{\mathrm{PL}}$ reduces to the constant $-2i\beta_0(\Delta-1)$.
	The stabilizer fixes the remaining representation data on this orbit, so the induced orbit representation is irreducible when no additional internal labels are present. \par

\subsection{3D Complete Carrollian Representation}\label{subsec:ThreeDimensionalRepresentation}

\subsubsection{Construction of the complete local representation}

	The 3D Carrollian conformal algebra is generated by $\{P^0,P^a,D,J^{12},B^a,K^0,K^a\}$, where $a=1,2$.
	With $\mu=0,1,2$ and $\epsilon^{12}=1$, its independent commutators are
	\begin{equation}\label{eq:ThreeDimensionalCarrollianAlgebra}
		\begin{aligned}
			&[D,P^\mu]=P^\mu,\qquad [D,K^\mu]=-K^\mu,\qquad [D,B^a]=[D,J^{12}]=0,\\
			&[J^{12},G^a]=\epsilon^{ab}G^b,\qquad G^a\in\{P^a,K^a,B^a\}, \qquad [J^{12},P^0]=[J^{12},K^0]=0,\\
			&[B^a,P^b]=\delta^{ab}P^0,\qquad [B^a,K^b]=\delta^{ab}K^0,\qquad [B^a,P^0]=[B^a,K^0]=0,\qquad [B^a,B^b]=0,\\
			&[P^0,K^a]=-2B^a,\qquad [P^a,K^0]=2B^a,\qquad [P^a,K^b]=-2\delta^{ab}D+2\epsilon^{ab}J^{12},\\
			&[P^0,P^a]=[P^a,P^b]=[K^0,K^a]=[K^a,K^b]=[P^0,K^0]=0.\\
		\end{aligned}
	\end{equation}
	As in the 2D construction, $[P^0,K^0]=0$ separates the descendants generated by temporal translations from the $K^0$ descendant direction included in our construction. \par

	To keep the rotation generator diagonal throughout the construction of the boost descendants, we introduce the complex combinations
	\begin{equation}
		B^\pm=\frac{1}{\sqrt{2}}(B^1\mp iB^2),\qquad K^\pm=\frac{1}{\sqrt{2}}(K^1\mp iK^2).
	\end{equation}
	They obey
	\begin{equation}
		[J^{12},B^\pm]=\pm iB^\pm,\qquad [J^{12},K^\pm]=\pm iK^\pm,\qquad [B^\pm,K^\mp]=K^0,\qquad [B^\pm,K^\pm]=0.
	\end{equation}
	Because $B^+$ and $B^-$ carry definite $J^{12}$ charges, the resulting discrete descendants can be chosen to be $J^{12}$ eigenoperators. \par

	We choose the rotation diagonal basis
	\begin{equation}
		\mathcal{O}_{\Delta,l}(n_{B^+},n_{B^-},n_{K^0}), \qquad n_{B^+},n_{B^-},n_{K^0}\in\mathbb{Z}_{\geq0},
	\end{equation}
	where an operator carrying any negative index is understood to vanish.
	The operator $\mathcal{O}_{\Delta,l}(0,0,0)$ is the Carrollian primary at the bottom of all three chains and satisfies
	\begin{equation}
		\begin{aligned}
			& [D,\mathcal{O}_{\Delta,l}(0,0,0)] =\Delta\mathcal{O}_{\Delta,l}(0,0,0),\qquad [J^{12},\mathcal{O}_{\Delta,l}(0,0,0)] =il\mathcal{O}_{\Delta,l}(0,0,0), \\
			& [B^\pm,\mathcal{O}_{\Delta,l}(0,0,0)] =[K^0,\mathcal{O}_{\Delta,l}(0,0,0)] =[K^\pm,\mathcal{O}_{\Delta,l}(0,0,0)] =0.
		\end{aligned}
	\end{equation}
	We normalize the lowering actions of $B^+$, $B^-$, and $K^0$ by
	\begin{equation}
		\begin{aligned}
			& [B^+,\mathcal{O}_{\Delta,l}(n_{B^+},n_{B^-},n_{K^0})] =\mathcal{O}_{\Delta,l}(n_{B^+}-1,n_{B^-},n_{K^0}), \\
			& [B^-,\mathcal{O}_{\Delta,l}(n_{B^+},n_{B^-},n_{K^0})] =\mathcal{O}_{\Delta,l}(n_{B^+},n_{B^-}-1,n_{K^0}), \\
			& [K^0,\mathcal{O}_{\Delta,l}(n_{B^+},n_{B^-},n_{K^0})] =\mathcal{O}_{\Delta,l}(n_{B^+},n_{B^-},n_{K^0}-1).
		\end{aligned}
	\end{equation}
	The algebra fixes the actions of the scaling, rotation, and $K^\pm$ generators to be
	\begin{equation}
		\begin{aligned}
			& [D,\mathcal{O}_{\Delta,l}(n_{B^+},n_{B^-},n_{K^0})] =(\Delta+n_{K^0}) \mathcal{O}_{\Delta,l}(n_{B^+},n_{B^-},n_{K^0}), \\
			& [J^{12},\mathcal{O}_{\Delta,l}(n_{B^+},n_{B^-},n_{K^0})] =i(l-n_{B^+}+n_{B^-}) \mathcal{O}_{\Delta,l}(n_{B^+},n_{B^-},n_{K^0}), \\
			& [K^+,\mathcal{O}_{\Delta,l}(n_{B^+},n_{B^-},n_{K^0})] =(n_{B^-}+c) \mathcal{O}_{\Delta,l}(n_{B^+},n_{B^-}+1,n_{K^0}-1), \\
			& [K^-,\mathcal{O}_{\Delta,l}(n_{B^+},n_{B^-},n_{K^0})] =(n_{B^+}+c) \mathcal{O}_{\Delta,l}(n_{B^+}+1,n_{B^-},n_{K^0}-1).
		\end{aligned}
	\end{equation}
	The commutators $[B^\mp,K^\pm]=K^0$ determine the dependence on $n_{B^\mp}$, and their action on the bottom of each boost chain fixes $c=1$.
	The displayed $J^{12}$ action makes explicit that every discrete descendant remains a rotation eigenoperator. \par

	Resumming the basis diagonal under rotations with the boost charges $\xi^+$ and $\xi^-$ gives
	\begin{equation}
		\begin{aligned}
			\mathcal{O}_{\Delta,l}(\xi^+,\xi^-,\kappa) &=\sum_{n_{B^+},n_{B^-},n_{K^0}=0}^{\infty} (\xi^+)^{n_{B^+}} (\xi^-)^{n_{B^-}} (i\kappa)^{n_{K^0}} \mathcal{O}_{\Delta,l}(n_{B^+},n_{B^-},n_{K^0}).
		\end{aligned}
	\end{equation}
	As in 2D, the discrete descendants are recovered by extracting the corresponding Taylor coefficients in $\xi^+$, $\xi^-$, and $\kappa$.
	The generating series contains only nonnegative powers of $\xi^+$, $\xi^-$, and $\kappa$.
	The factor of $i$ multiplying $\kappa$ is introduced for later convenience when imposing the reality conditions.
	In the $(\xi^+,\xi^-)$ basis, the actions at the origin are
	\begin{equation}
		\begin{aligned}
			& [B^\pm,\mathcal{O}_{\Delta,l}(\xi^+,\xi^-,\kappa)] =\xi^\pm\mathcal{O}_{\Delta,l}(\xi^+,\xi^-,\kappa),\\
			& [K^0,\mathcal{O}_{\Delta,l}(\xi^+,\xi^-,\kappa)] =i\kappa\mathcal{O}_{\Delta,l}(\xi^+,\xi^-,\kappa), \\
			& [K^\pm,\mathcal{O}_{\Delta,l}(\xi^+,\xi^-,\kappa)] =i\kappa\partial_{\xi^\mp} \mathcal{O}_{\Delta,l}(\xi^+,\xi^-,\kappa), \\
			& [D,\mathcal{O}_{\Delta,l}(\xi^+,\xi^-,\kappa)] =(\kappa\partial_\kappa+\Delta) \mathcal{O}_{\Delta,l}(\xi^+,\xi^-,\kappa), \\
			& [J^{12},\mathcal{O}_{\Delta,l}(\xi^+,\xi^-,\kappa)] =i(l-\xi^+\partial_{\xi^+}+\xi^-\partial_{\xi^-})\mathcal{O}_{\Delta,l}(\xi^+,\xi^-,\kappa).
		\end{aligned}
	\end{equation}
	\par

\begin{figure}
		\centering
		\captionsetup{width=0.8\linewidth}
		\includegraphics[width=0.4\linewidth]{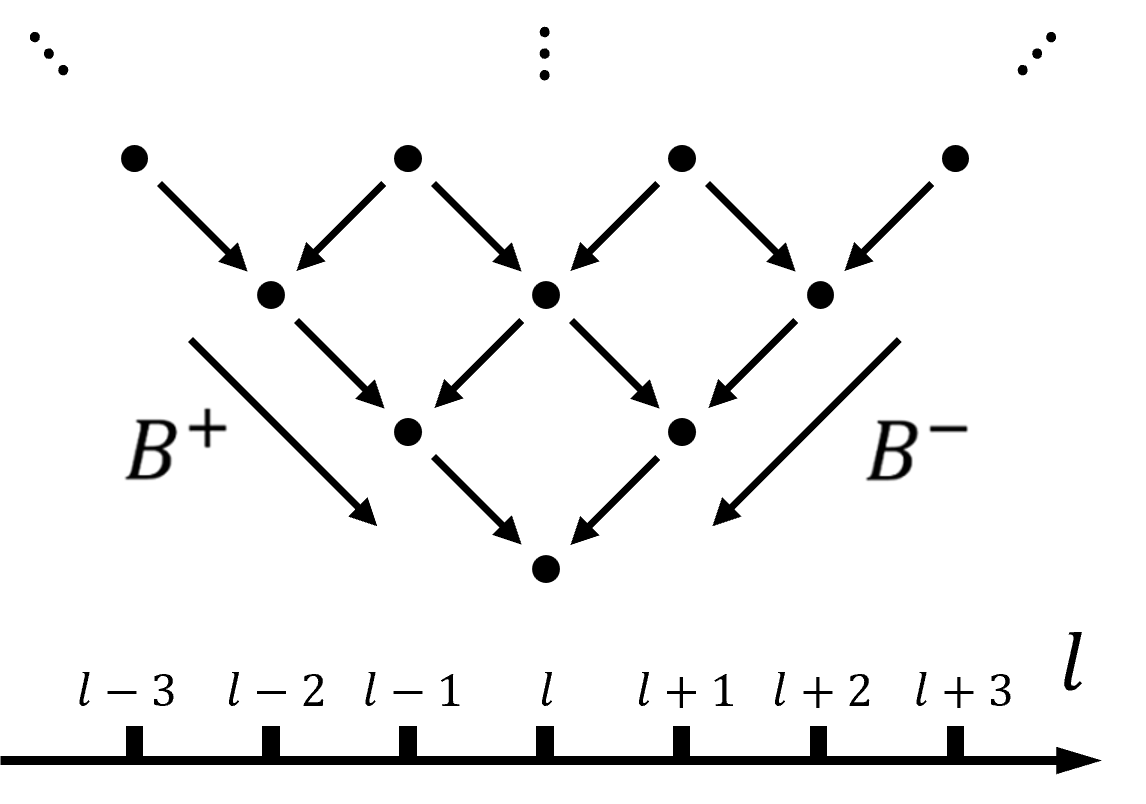}
		\caption{Schematic infinite dimensional multiplets of the 3D Carrollian rotation subalgebra, following \cite{Chen:2021xkw}.
			Each node represents a $J^{12}$ eigenoperator, and the $B^+$ and $B^-$ actions connect components with neighboring rotation weights.}
		\label{fig:ThreeDimensionalISO2Representations}
	\end{figure}

	The real boost charge labels are then introduced through
	\begin{equation}
		\xi^\pm=\frac{1}{\sqrt{2}}(i\beta^1\pm\beta^2).
	\end{equation}
	The generating operator in the real basis is defined by
	\begin{equation}
		\mathcal{O}_{\Delta,l}(\beta^a,\kappa) \equiv\mathcal{O}_{\Delta,l} (\xi^+(\beta^a),\xi^-(\beta^a),\kappa).
	\end{equation}
	The rotation action then becomes $\beta^1\partial_{\beta^2}-\beta^2\partial_{\beta^1}+il$, in agreement with the $J^{12}$ eigenvalues of the discrete descendants.
	The actions at the origin are
	\begin{equation}
		\begin{aligned}
			& [B^a,\mathcal{O}_{\Delta,l}(\beta^a,\kappa)]=i\beta^a\mathcal{O}_{\Delta,l}(\beta^a,\kappa),\\
			& [K^0,\mathcal{O}_{\Delta,l}(\beta^a,\kappa)]=i\kappa\mathcal{O}_{\Delta,l}(\beta^a,\kappa), \\
			& [K^a,\mathcal{O}_{\Delta,l}(\beta^a,\kappa)]=\kappa\partial_{\beta^a}\mathcal{O}_{\Delta,l}(\beta^a,\kappa), \\
			& [J^{12},\mathcal{O}_{\Delta,l}(\beta^a,\kappa)]=(\beta^1\partial_{\beta^2}-\beta^2\partial_{\beta^1}+il)\mathcal{O}_{\Delta,l}(\beta^a,\kappa).
		\end{aligned}
	\end{equation}
	The infinite dimensional representation structure of the Carrollian rotation subalgebra was analyzed in \cite{Chen:2021xkw}.
	In two spatial dimensions, the subalgebra generated by $J^{12}$, $B^1$, and $B^2$ is isomorphic to $\mathfrak{iso}(2)$.
	The generators $B^+$ and $B^-$ connect neighboring rotation weights in the infinite dimensional multiplets illustrated in Figure \ref{fig:ThreeDimensionalISO2Representations}.
	Allowing arbitrary nonnegative $n_{B^+}$ and $n_{B^-}$ produces an infinite dimensional representation.
	The continuous boost charges $\beta^a$ organize this representation and transform covariantly under rotations, while every term in the discrete expansion remains a $J^{12}$ eigenoperator. \par

	The hypersurface $\kappa=0$ carries the boost charge representation organized by the labels $\beta^a$, which transform covariantly under rotations, whereas the complete family includes the independent $K^0$ descendant direction.
	Unlike the 2D label $\beta$, the vector $\beta^a$ is not invariant on this hypersurface.
	Rather, $\vec{\beta}^{\,2}$ remains fixed. \par

	A local operator is obtained by defining
	\begin{equation}
		U(u,z^a)=\exp(uP^0+z^aP^a),
	\end{equation}
	and translating the operator in the complete representation according to
	\begin{equation}
		\mathcal{O}_{\Delta,l}(u,z^a;\beta^a,\kappa) =U(u,z^a)\mathcal{O}_{\Delta,l}(\beta^a,\kappa)U^{-1}(u,z^a).
	\end{equation}
	For the spin term in the local representation, we use $\epsilon^{12}=1$.
	The Baker--Campbell--Hausdorff formula gives the complete local differential representation
	\begin{equation}\label{eq:ThreeDimensionalRepresentation}
		\begin{aligned}
			& [P^0,\mathcal{O}_{\Delta,l}(u,z^a;\beta^a,\kappa)] =\partial_u\mathcal{O}_{\Delta,l}(u,z^a;\beta^a,\kappa), \\
			& [P^a,\mathcal{O}_{\Delta,l}(u,z^a;\beta^a,\kappa)] =\partial_{z^a}\mathcal{O}_{\Delta,l}(u,z^a;\beta^a,\kappa), \\
			& [D,\mathcal{O}_{\Delta,l}(u,z^a;\beta^a,\kappa)] =(u\partial_u+z^a\partial_{z^a}+\kappa\partial_\kappa+\Delta) \mathcal{O}_{\Delta,l}(u,z^a;\beta^a,\kappa), \\
			& [J^{12},\mathcal{O}_{\Delta,l}(u,z^a;\beta^a,\kappa)] =(z^1\partial_{z^2}-z^2\partial_{z^1} +\beta^1\partial_{\beta^2}-\beta^2\partial_{\beta^1}+il) \mathcal{O}_{\Delta,l}(u,z^a;\beta^a,\kappa), \\
			& [B^a,\mathcal{O}_{\Delta,l}(u,z^a;\beta^a,\kappa)] =(z^a\partial_u+i\beta^a) \mathcal{O}_{\Delta,l}(u,z^a;\beta^a,\kappa), \\
			& [K^0,\mathcal{O}_{\Delta,l}(u,z^a;\beta^a,\kappa)] =(-\vec{z}^{\,2}\partial_u-2i\vec{z}\cdot\vec{\beta}+i\kappa) \mathcal{O}_{\Delta,l}(u,z^a;\beta^a,\kappa), \\
			& [K^a,\mathcal{O}_{\Delta,l}(u,z^a;\beta^a,\kappa)] =\left( 2uz^a\partial_u+2z^az^b\partial_{z^b}-\vec{z}^{\,2}\partial_{z^a} +2z^a(\kappa\partial_\kappa+\Delta) \right. \\
			& \qquad\qquad\qquad\qquad\left. +2z^b(\beta^a\partial_{\beta^b}-\beta^b\partial_{\beta^a} +il\epsilon^{ab}) +\kappa\partial_{\beta^a}+2ui\beta^a \right) \mathcal{O}_{\Delta,l}(u,z^a;\beta^a,\kappa).
		\end{aligned}
	\end{equation}
	The differential operators in \eqref{eq:ThreeDimensionalRepresentation} close on the algebra \eqref{eq:ThreeDimensionalCarrollianAlgebra}.
	The complete local representation therefore extends the discrete multiplets diagonal under rotations by the continuous boost charge labels $\beta^a$ and the independent descendant label $\kappa$ associated with $K^0$. \par

\subsubsection{Properties of the complete representation}

	This subsection examines the reality and duality properties of the complete 3D representation before turning to its representation orbits and Casimir structure. \par

	We retain the anti-Hermitian convention $G^\dagger=-G$ and the conjugation rule established in Subsection \ref{subsec:TwoDimensionalCompletedRepresentationProperties}.
	When $\kappa$ is an independent representation label, the corresponding pairing in 3D is
	\begin{equation}
		\langle f,g\rangle =\int du\,d^2z\,d^2\beta\,d\kappa\, f^*(u,z^a;\beta^a,\kappa) g(u,z^a;\beta^a,\kappa),
	\end{equation}
	where boundary terms are assumed to vanish.
	For the differential realization of $D$ in \eqref{eq:ThreeDimensionalRepresentation}, integration by parts gives
	\begin{equation}
		\left\langle f,(u\partial_u+z^a\partial_{z^a}+\kappa\partial_\kappa)g\right\rangle=-\left\langle(u\partial_u+z^a\partial_{z^a}+\kappa\partial_\kappa+4)f,g\right\rangle.
	\end{equation}
	Here the constant $4$ is the divergence $\partial_u u+\partial_{z^a}z^a+\partial_\kappa\kappa$, with one contribution from $u$, two from $z^a$, and one from $\kappa$.
	The condition $D^\dagger=-D$ therefore requires
	\begin{equation}
		\Delta^\dagger=4-\Delta.
	\end{equation}
	The resulting conformal dimension lies on the principal line,
	\begin{equation}
		\Delta=2+i\nu, \qquad \nu\in\mathbb{R}.
	\end{equation}
	Applying the same anti-Hermitian convention to the $B^a$, $K^0$, and $J^{12}$ actions gives
	\begin{equation}
		(\beta^a)^\dagger=\beta^a, \qquad \kappa^\dagger=\kappa, \qquad l^\dagger=l.
	\end{equation}
	Thus $\beta^a$, $\kappa$, and $l$ are real, while the conformal dimension lies on the principal line.
	We take the global rotation group to be $\mathrm{Spin}(2)$, so that $l\in\frac{1}{2}\mathbb{Z}$. \par

	Using $([G,\mathcal{O}])^\dagger=[G,\mathcal{O}^\dagger]$, the defining actions at the origin become
	\begin{equation}
		\begin{aligned}
			& [B^a,\mathcal{O}_{\Delta,l}^\dagger(\beta^a,\kappa)] =-i\beta^a\mathcal{O}_{\Delta,l}^\dagger(\beta^a,\kappa),\\
			& [K^0,\mathcal{O}_{\Delta,l}^\dagger(\beta^a,\kappa)] =-i\kappa\mathcal{O}_{\Delta,l}^\dagger(\beta^a,\kappa), \\
			& [J^{12},\mathcal{O}_{\Delta,l}^\dagger(\beta^a,\kappa)] =(-il+\beta^1\partial_{\beta^2}-\beta^2\partial_{\beta^1})\mathcal{O}_{\Delta,l}^\dagger(\beta^a,\kappa).
		\end{aligned}
	\end{equation}
	The conjugated operator therefore carries the labels $-\beta^a$, $-\kappa$, and $-l$.
	Denoting the dual operator by $\mathcal{O}_{4-\Delta,-l}^\prime$, we obtain
	\begin{equation}\label{eq:ThreeDimensionalDualOperator}
		\mathcal{O}_{\Delta,l}^\dagger(u,z^a;\beta^a,\kappa) =\mathcal{O}_{4-\Delta,-l}^\prime(u,z^a;-\beta^a,-\kappa).
	\end{equation}
	On the invariant $\kappa=0$ sector, we instead use
	\begin{equation}
		\langle f,g\rangle_{\kappa=0}=\int du\,d^2z\,d^2\beta\,f^*(u,z^a;\beta^a,0)g(u,z^a;\beta^a,0).
	\end{equation}
	The dual dimension and allowed line on this sector are
	\begin{equation}
		\Delta=\frac{3}{2}+i\nu, \qquad \nu\in\mathbb{R}.
	\end{equation}
	The conjugated operator in the $\kappa=0$ sector is therefore
	\begin{equation}\label{eq:ThreeDimensionalZeroKappaDualOperator}
		\mathcal{O}_{\Delta,l}^{\dagger}(u,z^a;\beta^a,0)=\mathcal{O}_{3-\Delta,-l}^{\prime}(u,z^a;-\beta^a,0).
	\end{equation}
	The same interpretation of the state space applies in 3D.
	For $\kappa\neq0$, the contact two-point functions in Subsection \ref{subsec:ThreeDimensionalGenericKappaCorrelators} normalize the states through Dirac distributions supported at a point.
	On the enlarged blow-up configuration space, $\delta(r_z)$ retains the angular data on the circle at $\lvert\vec{z}\rvert=0$, while the boost label support remains an ordinary vector distribution.
	In the $\kappa=0$ sector with $\vec{\beta}\neq0$, the two-point function between an operator and its dual vanishes, so the corresponding state is null.
	At $\vec{\beta}=\kappa=0$, both electric contact pairing and magnetic non-contact pairing exist.
	Thus the existence of an algebraic dual representation does not by itself guarantee that the vacuum two-point pairing is nonzero.
	The correlation functions distinguish the $\kappa\neq0$ orbits normalized by delta functions, the $\kappa=0$ orbits with nonzero boost labels and zero norm, and the special orbit with $\vec{\beta}=\kappa=0$ that also admits a non-contact pairing. \par

	\begin{table}[htpb]
		\renewcommand{\arraystretch}{2}
		\centering
		\caption{Finite transformations of the representation variables in 3D.
			The parameters $u_0$, $z_0^a$, $\lambda>0$, $\theta$, $b^a$, $k^0$, and $k^a$ label the corresponding transformations.
			The last column contains the Fourier variable $\rho$, conjugate to $u$, which is introduced in \eqref{eq:ThreeDimensionalFourierTransform}.}
		\label{tab:ThreeDimensionalCompletedFiniteTransformations}
		\resizebox{\linewidth}{!}{
			\begin{tabular}{c|c|c|c|c|c}
				\hline
				Generator & $u'$ & $z^{\prime a}$ & $\beta^{\prime a}$ & $\kappa'$ & $\rho'$ \\\hline\hline
				$P^0$
					& $u+u_0$
					& $z^a$
					& $\beta^a$
					& $\kappa$
					& $\rho$ \\\hline
				$P^a$
					& $u$
					& $z^a+z_0^a$
					& $\beta^a$
					& $\kappa$
					& $\rho$ \\\hline
				$D$
					& $\lambda u$
					& $\lambda z^a$
					& $\beta^a$
					& $\lambda\kappa$
					& $\lambda^{-1}\rho$ \\\hline
				$J^{12}$
					& $u$
					& $\begin{pmatrix}
						\cos(\theta)z^1-\sin(\theta)z^2 \\
						\sin(\theta)z^1+\cos(\theta)z^2
					\end{pmatrix}$
					& $\begin{pmatrix}
						\cos(\theta)\beta^1-\sin(\theta)\beta^2 \\
						\sin(\theta)\beta^1+\cos(\theta)\beta^2
					\end{pmatrix}$
					& $\kappa$
					& $\rho$ \\\hline
				$B^a$
					& $u+\vec{b}\cdot\vec{z}$
					& $z^a$
					& $\beta^a$
					& $\kappa$
					& $\rho$ \\\hline
				$K^0$
					& $u-k^0\vec{z}^{\,2}$
					& $z^a$
					& $\beta^a$
					& $\kappa$
					& $\rho$ \\\hline
				$K^a$
					& $\dfrac{u}{1-2\vec{k}\cdot\vec{z}+\vec{k}^{\,2}\vec{z}^{\,2}}$
					& $\dfrac{z^a-k^a\vec{z}^{\,2}}
						{1-2\vec{k}\cdot\vec{z}+\vec{k}^{\,2}\vec{z}^{\,2}}$
					& $\begin{aligned}
						&\frac{\left((1-\vec{k}\cdot\vec{z})^2-(\epsilon^{bc}k^bz^c)^2\right)\beta^a}{1-2\vec{k}\cdot\vec{z}+\vec{k}^{\,2}\vec{z}^{\,2}}\\
						&\qquad - \frac{2(1-\vec{k}\cdot\vec{z})
						\left(k^a(\vec{z}\cdot\vec{\beta})-z^a(\vec{k}\cdot\vec{\beta})\right)}{1-2\vec{k}\cdot\vec{z}+\vec{k}^{\,2}\vec{z}^{\,2}}\\
						& \qquad + \frac{\left(k^a(1-2\vec{k}\cdot\vec{z})+\vec{k}^{\,2}z^a\right)\kappa}{1-2\vec{k}\cdot\vec{z}+\vec{k}^{\,2}\vec{z}^{\,2}}
					\end{aligned}$
					& $\dfrac{\kappa}{1-2\vec{k}\cdot\vec{z}+\vec{k}^{\,2}\vec{z}^{\,2}}$
					& $\begin{aligned}
						& (1-2\vec{k}\cdot\vec{z}+\vec{k}^{\,2}\vec{z}^{\,2})\rho \\
						& \quad
						+2(\vec{k}-\vec{k}^{\,2}\vec{z})\cdot\vec{\beta}
						+\vec{k}^{\,2}\kappa
					\end{aligned}$ \\\hline
			\end{tabular}
		}
	\end{table}

	Exponentiating the vector field parts of \eqref{eq:ThreeDimensionalRepresentation} gives the finite transformations shown in Table \ref{tab:ThreeDimensionalCompletedFiniteTransformations}.
	For a regular finite $K^a$ transformation, the common denominator obeys
	\begin{equation}
		1-2\vec{k}\cdot\vec{z}+\vec{k}^{\,2}\vec{z}^{\,2} =(1-\vec{k}\cdot\vec{z})^2+(\epsilon^{ab}k^az^b)^2 >0.
	\end{equation}
	It follows that the sign of $\kappa$ cannot change under a regular transformation connected to the identity.
	When $\kappa=0$, the transformation of $\beta^a$ is an $SO(2)$ rotation and preserves $\vec{\beta}^{\,2}$.
	For fixed $\Delta$ and $l$, these finite transformations determine the representation orbits.
	When $\kappa\neq0$, translations first set $u=z^a=0$, a $K^a$ transformation with $k^a=-\beta^a/\kappa$ then sets $\beta^a=0$, and a dilatation fixes $\kappa$ to its sign.
	When $\kappa=0$, translations set $u=z^a=0$, and rotations connect all $\beta^a$ with the same $\vec{\beta}^{\,2}$.
	The representation orbits therefore have the representatives
	\begin{equation}
		\begin{aligned}
			& \kappa>0: && (u,z^a;\beta^a,\kappa)=(0,0;0,1), \\
			& \kappa<0: && (u,z^a;\beta^a,\kappa)=(0,0;0,-1), \\
			& \kappa=0,\quad \vec{\beta}^{\,2}=r_{\beta}^2: && (u,z^a;\beta^a,\kappa)=(0,0;(r_{\beta},0),0), \qquad r_{\beta}\geq0.
		\end{aligned}
	\end{equation}
	The duality map exchanges the $\kappa>0$ and $\kappa<0$ orbits.
	On a $\kappa=0$ orbit, $\beta^a\mapsto-\beta^a$ is a rotation by $\pi$, so the representation orbit is unchanged as a set, while the representation label is mapped from $l$ to $-l$. \par

	We define the operator after Fourier transformation using the same convention as in \eqref{eq:TwoDimensionalFourierTransform}.
	\begin{equation}\label{eq:ThreeDimensionalFourierTransform}
		\mathcal{O}_{\Delta,l}(u,z^a;\beta^a,\kappa) =\int\frac{d\rho}{\sqrt{2\pi}}\, e^{-i\rho u} \widetilde{\mathcal{O}}_{\Delta,l}(\rho,z^a;\beta^a,\kappa).
	\end{equation}
	Here $\rho\in\mathbb{R}$, and
	\begin{equation}
		\partial_u\longmapsto-i\rho, \qquad u\longmapsto-i\partial_\rho, \qquad u\partial_u\longmapsto-\left(\rho\partial_\rho+1\right).
	\end{equation}
	The Fourier transform preserves the reality conditions $\Delta=2+i\nu$ and $\rho,\beta^a,\kappa,l,\nu\in\mathbb{R}$.
	Hermitian conjugation reverses $\rho$, so the momentum space dual operator satisfies
	\begin{equation}
		\widetilde{\mathcal{O}}_{\Delta,l}^\dagger(\rho,z^a;\beta^a,\kappa) =\widetilde{\mathcal{O}}_{4-\Delta,-l}^\prime(-\rho,z^a;-\beta^a,-\kappa).
	\end{equation}
	The nontrivial momentum space actions of $D$ and $K^a$ are
	\begin{equation}
		\begin{aligned}
			& [D,\widetilde{\mathcal{O}}_{\Delta,l}(\rho,z^a;\beta^a,\kappa)] =(-\rho\partial_\rho+z^a\partial_{z^a}+\kappa\partial_\kappa+\Delta-1)\widetilde{\mathcal{O}}_{\Delta,l}(\rho,z^a;\beta^a,\kappa), \\
			& [K^a,\widetilde{\mathcal{O}}_{\Delta,l}(\rho,z^a;\beta^a,\kappa)] =\left( 2z^az^b\partial_{z^b}-\vec{z}^{\,2}\partial_{z^a} +2z^a\kappa\partial_\kappa +2z^b(\beta^a\partial_{\beta^b}-\beta^b\partial_{\beta^a}) \right. \\
			& \qquad\qquad\qquad\qquad\left. +\kappa\partial_{\beta^a} +2(\beta^a-z^a\rho)\partial_\rho +2z^a(\Delta-1) +2il\epsilon^{ab}z^b \right) \widetilde{\mathcal{O}}_{\Delta,l}(\rho,z^a;\beta^a,\kappa).
		\end{aligned}
	\end{equation}
	In particular, $2uz^a\partial_u$ and $2ui\beta^a$ respectively produce $-2z^a(\rho\partial_\rho+1)$ and $2\beta^a\partial_\rho$, and both contributions are required for the transformation of $\rho$ listed in Table \ref{tab:ThreeDimensionalCompletedFiniteTransformations}. \par

	The quadratic Casimir of the 3D Carrollian conformal algebra is
	\begin{equation}\label{eq:3DQuadraticCasimir}
		\mathcal{C}_2 =(B^1)^2+(B^2)^2+P^0K^0 =i\kappa\partial_u-\vec{\beta}^{\,2} \xrightarrow{\ \mathcal{F}_u\ } \rho\kappa-\vec{\beta}^{\,2}.
	\end{equation}
	Its invariance under translations, boosts, $K^0$, and rotations is immediate.
	Under a finite dilatation,
	\begin{equation}
		(\lambda^{-1}\rho)(\lambda\kappa)-\vec{\beta}^{\,2} =\rho\kappa-\vec{\beta}^{\,2}.
	\end{equation}
	For the finite $K^a$ transformation in Table \ref{tab:ThreeDimensionalCompletedFiniteTransformations}, direct substitution gives
	\begin{equation}
		\begin{aligned}
			(\vec{\beta}^{\,\prime})^2 &=\vec{\beta}^{\,2} +\frac{2\kappa(\vec{k}-\vec{k}^{\,2}\vec{z})\cdot\vec{\beta}+\vec{k}^{\,2}\kappa^2} {1-2\vec{k}\cdot\vec{z}+\vec{k}^{\,2}\vec{z}^{\,2}}, \qquad
			\rho'\kappa' &=\rho\kappa +\frac{2\kappa(\vec{k}-\vec{k}^{\,2}\vec{z})\cdot\vec{\beta}+\vec{k}^{\,2}\kappa^2} {1-2\vec{k}\cdot\vec{z}+\vec{k}^{\,2}\vec{z}^{\,2}}.
		\end{aligned}
	\end{equation}
	Subtracting these two expressions gives
	\begin{equation}
		\rho'\kappa'-(\vec{\beta}^{\,\prime})^2 =\rho\kappa-\vec{\beta}^{\,2}.
	\end{equation}
	Thus $\mathcal{C}_2$ is invariant under all finite Carrollian conformal transformations.
	In the realization in Fourier space, $\mathcal{C}_2$ acts multiplicatively, so the representation space first decomposes into sectors of fixed $\mathcal{C}_2$. \par

	The algebra also admits a quartic Casimir.
	In the differential realization \eqref{eq:ThreeDimensionalRepresentation}, it acts as
	\begin{equation}\label{eq:3DQuarticCasimir}
		\begin{aligned}
			\mathcal{C}_4={}&-(\Delta-2)(\Delta-1)\vec{\beta}^{\,2}
			-2\kappa(\Delta-1)\vec{\beta}^{\,2}\partial_{\kappa} \\
			&+\kappa\left((1-\Delta)\beta^1+il\beta^2\right)\partial_{z^1}
			+\kappa\left((1-\Delta)\beta^2-il\beta^1\right)\partial_{z^2}
			+i\kappa\left(l^2+1-\Delta\right)\partial_u \\
			&-\kappa^2\vec{\beta}^{\,2}\partial_{\kappa}^2
			-\kappa^2\beta^1\partial_{z^1}\partial_{\kappa}
			-\kappa^2\beta^2\partial_{z^2}\partial_{\kappa}
			-\frac{\kappa^2}{4}\left(\partial_{z^1}^2+\partial_{z^2}^2\right)
			-i\kappa^2\partial_u\partial_{\kappa} \\
			&+\kappa\left(i(1-\Delta)\beta^1-l\beta^2\right)\partial_u\partial_{\beta^1}
			+\kappa\left(l\beta^1+i(1-\Delta)\beta^2\right)\partial_u\partial_{\beta^2} \\
			&-i\kappa^2\beta^1\partial_u\partial_{\beta^1}\partial_{\kappa}
			-i\kappa^2\beta^2\partial_u\partial_{\beta^2}\partial_{\kappa}
			-\frac{i\kappa^2}{2}\left(\partial_u\partial_{z^1}\partial_{\beta^1}
			+\partial_u\partial_{z^2}\partial_{\beta^2}\right) \\
			&+\frac{\kappa^2}{4}\partial_u^2
			\left(\partial_{\beta^1}^2+\partial_{\beta^2}^2\right).
		\end{aligned}
	\end{equation}
	This operator commutes with every generator in eq. \eqref{eq:ThreeDimensionalCarrollianAlgebra}.
	For $\kappa\neq0$, it contains derivatives and is not proportional to the identity, so the corresponding orbit representations are reducible.
	Each sector with fixed $\mathcal{C}_2$ can be decomposed further by the action of $\mathcal{C}_4$.
	At $\kappa=0$, it reduces to
	\begin{equation}
		\left.\mathcal{C}_4\right|_{\kappa=0}
		=-(\Delta-2)(\Delta-1)\vec{\beta}^{\,2}.
	\end{equation}
	Both Casimirs are then constant on an orbit of fixed $\vec{\beta}^{\,2}$.
	The stabilizer fixes the remaining representation data, so the induced orbit representation is irreducible when no additional internal labels are present. \par

	The 3D Carrollian representations commonly used in holographic constructions in flat space do not introduce an independent boost charge label \cite{Bagchi:2016bcd,Chen:2021xkw,Salzer:2023jqv}.
	In our notation, these representations lie at $\vec{\beta}=0$ and $\kappa=0$ and therefore have $\mathcal{C}_2=0$, whereas the complete representation contains invariant sets with $\mathcal{C}_2\neq0$.
	For real $\rho$ and $\vec{\beta}$, positive $m^2$ can be represented by
	\begin{equation}\label{eq:MassiveCasimirRelation}
		m^2=\mathcal{C}_2=\kappa\rho-\vec{\beta}^{\,2},\qquad
		\rho=\frac{m^2+\vec{\beta}^{\,2}}{\kappa}.
	\end{equation}
	At $\kappa=0$, the first equality gives $m^2=-\vec{\beta}^{\,2}\leq0$, which has no real solution for $m^2>0$.
	For $\kappa\neq0$, the second equality gives a real solution.
	The sets with nonzero $\mathcal{C}_2$ are therefore natural candidates for a massive Carrollian dictionary.
	The explicit construction of this dictionary is discussed in \cite{workinprogress}.
	The next section examines this representation structure through Ward identities and correlation functions between different representation orbits. \par

\section{Two-Point Correlation Functions}\label{sec:CorrelationFunctions}

	To examine how the additional orbit labels enter symmetry constraints, we calculate the corresponding two-point functions.
	We use global Ward identities for the 2D and 3D representations constructed in Section \ref{sec:CompleteCarrollianRepresentations}.
	The analysis treats separately the three possible orbit configurations distinguished by whether each insertion has $\kappa=0$, because their invariant variables and distributional supports differ. \par

	For an arbitrary global symmetry generator $G$, we define the two-point correlation function by
	\begin{equation}
		\mathcal{G}\equiv\bra{0}\mathcal{O}_1\mathcal{O}_2\ket{0}.
	\end{equation}
	Invariance of the vacuum requires
	\begin{equation}
		G\ket{0}=0,\qquad \bra{0}G=0.
	\end{equation}
	The resulting Ward identity is
	\begin{equation}
		0=\bra{0}[G,\mathcal{O}_1\mathcal{O}_2]\ket{0} =\expval{[G,\mathcal{O}_1]\mathcal{O}_2}+\expval{\mathcal{O}_1[G,\mathcal{O}_2]} =\left(\mathcal{D}_{G,1}+\mathcal{D}_{G,2}\right)\mathcal{G}.
	\end{equation}
	Here $\mathcal{D}_{G,i}$ acts on both the Carrollian coordinates and the representation variables of the $i$th insertion, so the same Ward identities constrain spacetime dependence, orbit labels, and spin data. \par

	We first impose the translation Ward identities
	\begin{equation}
		\left(\partial_{x_1^\mu}+\partial_{x_2^\mu}\right)\mathcal{G}=0.
	\end{equation}
	They restrict the spacetime dependence to coordinate differences.
	Accordingly, we use $u=u_1-u_2$ together with $z=z_1-z_2$ in Subsection \ref{subsec:TwoDimensionalCorrelators}, and with $z^a=z_1^a-z_2^a$ in Subsection \ref{subsec:ThreeDimensionalCorrelators}.
	All remaining Ward identities are written in these variables and act simultaneously on the two sets of representation labels. \par

	Within a fixed orbit configuration, the Ward identities determine the distributional support, the selection rules, and the kinematic factors.
	They also identify the invariant combinations that can carry dynamical information.
	The candidate invariants $\mathcal{I}_A$ are invariant under the Carrollian transformations listed in Table \ref{tab:TwoDimensionalCompletedFiniteTransformations} and Table \ref{tab:ThreeDimensionalCompletedFiniteTransformations}.
	A solution therefore has the schematic form
	\begin{equation}
		\mathcal{G}=f(\mathcal{I}_A)\,\mathcal{K}\,\mathcal{S}.
	\end{equation}
	The function $f$ is arbitrary in the invariant arguments.
	The factor $\mathcal{K}$ contains the kinematic dependence fixed by symmetry, while $\mathcal{S}$ specifies the distributional support.
	Unlike an ordinary CFT two-point function, which is fixed by global conformal symmetry up to an overall constant, the Carrollian solutions obtained here can retain an arbitrary function of the invariant arguments.
	Selection rules for the external representation labels are stated separately.
	We call a solution non-contact when its support permits spatially separated insertions and contact when it is supported at coincident spatial points. \par

	Subsection \ref{subsec:TwoDimensionalCorrelators} first solves the 2D problem and establishes the common organization of the three orbit configurations.
	Subsection \ref{subsec:ThreeDimensionalCorrelators} then performs the 3D analysis, including the angular and spin data that arise at contact.
	For each configuration, we identify the candidate invariants, impose the distributional constraints, solve separately for magnetic non-contact and electric contact branches, and finally specialize to an operator and its conjugate whenever the dual representation belongs to the same configuration.
	The resulting correlators distinguish the orbit configurations because they have different supports and cannot be obtained from one another by simply setting representation labels to zero. \par

\subsection{2D Carrollian Two-Point Correlation Functions}\label{subsec:TwoDimensionalCorrelators}

	We classify the 2D correlators according to whether both $\kappa_i$ are nonzero, both vanish, or only one is nonzero.
	The three orbit configurations are treated in Subsections \ref{subsec:TwoDimensionalGenericKappaCorrelators}, \ref{subsec:TwoDimensionalZeroKappaCorrelators}, and \ref{subsec:TwoDimensionalMixedKappaCorrelators}.
	For the first two configurations, we present the non-contact solution, determine the independent contact solution, and then impose the conditions for conjugate representations.
	The mixed configuration admits only a non-contact solution.
	% The resulting 2D correlators are important in their own right and also provide the organizational template for the 3D analysis. \par

	The Ward identities used below follow directly from the differential representation \eqref{eq:TwoDimensionalRepresentation}.
	Since this representation has already been established, we do not repeat the full set of Ward identities in every orbit configuration.
	We instead summarize the consequences by listing the candidate invariants, the independent arguments of $f$ on each branch, the kinematic factors, the distributional support, and the selection rules. \par

\subsubsection{Both insertions with \texorpdfstring{$\kappa_1\neq0$}{kappa1 nonzero} and \texorpdfstring{$\kappa_2\neq0$}{kappa2 nonzero}}\label{subsec:TwoDimensionalGenericKappaCorrelators}

	We first consider the non-contact solution for a general pair of operators in $\kappa_i\neq0$ orbits, without imposing additional conditions on their representation parameters.
	The two insertions admit three candidate invariants.
	\begin{equation}\label{eq:2DGenericInvariantCombinations}
		\begin{aligned}
			\mathcal{I}_1&=\frac{z^2}{\kappa_1\kappa_2},\qquad \mathcal{I}_2=\frac{z(\beta_1+\beta_2)-\kappa_1+\kappa_2}{z},\qquad
			\mathcal{I}_3&=\frac{\beta_2(z\beta_1-\kappa_1)+\beta_1\kappa_2}{z}.
		\end{aligned}
	\end{equation}
	Unlike an ordinary conformal two-point function, for which no conformal cross ratio exists, the complete Carrollian representation admits nontrivial invariant dependence already for two insertions.
	After imposing the distributional support, the combinations that remain independent become the arguments of the arbitrary function $f$ in the two-point correlator. \par

	The Ward identities of $B^1$ and $K^0$ fix the dependence on the time separation to be
	\begin{equation}
		\exp\!\left( -\frac{iu\left(\beta_1+\beta_2\right)}{z} \right),
	\end{equation}
	and impose the distributional support
	\begin{equation}
		\delta\!\left( z-\frac{\kappa_1+\kappa_2}{\beta_1-\beta_2} \right).
	\end{equation}
	On this support, only two independent combinations from \eqref{eq:2DGenericInvariantCombinations} remain.
	\begin{equation}
		\frac{z^2}{\kappa_1\kappa_2}, \qquad \frac{ \kappa_1\beta_2^2+\kappa_2\beta_1^2 }{ \kappa_1+\kappa_2 }.
	\end{equation}
	They therefore provide the natural arguments of the dynamical function $f$. \par

	The resulting non-contact two-point function is
	\begin{equation}\label{eq:TwoDimensionalGenericCorrelator}
		\begin{aligned}
			& \mathcal{G}_{\mathrm{m}} = f\!\left( \frac{z^2}{\kappa_1\kappa_2}, \frac{ \kappa_1\beta_2^2+\kappa_2\beta_1^2 }{ \kappa_1+\kappa_2 } \right) \exp\!\left( -\frac{iu\left(\beta_1+\beta_2\right)}{z} \right) \frac{ \abs{\kappa_1}^{\frac{1}{2}-\Delta_1} \abs{\kappa_2}^{\frac{1}{2}-\Delta_2} }{ \abs{\beta_1-\beta_2} } \delta\!\left( z-\frac{\kappa_1+\kappa_2}{\beta_1-\beta_2} \right).
		\end{aligned}
	\end{equation}
	Several features of this correlator deserve emphasis.
	There is no selection rule relating $\Delta_1$ and $\Delta_2$ for the general non-contact solution.
	The $\kappa_i$ factors together have scaling degree $1-\Delta_1-\Delta_2$, while the shifted Dirac distribution has scaling degree $-1$, so their product has the required degree $-\Delta_1-\Delta_2$.
	Because the real parameters $\kappa_i$ may have either sign and the exponents need not be integers, the homogeneous factors are written with absolute values.
	The apparent singularity in $1/\abs{\beta_1-\beta_2}$ is not a genuine singularity.
	It is solely the Jacobian produced by rewriting the Dirac distribution in shifted form.
	An equivalent form of the distributional factor is
	\begin{equation}
		\frac{1}{\abs{\beta_1-\beta_2}} \delta\!\left( z-\frac{\kappa_1+\kappa_2}{\beta_1-\beta_2} \right) = \delta\!\left( z\left(\beta_1-\beta_2\right)-\kappa_1-\kappa_2 \right).
	\end{equation}
	The arbitrary function $f$ contains the dynamical information through its dependence on these two independent combinations.
	We call \eqref{eq:TwoDimensionalGenericCorrelator} a generalized magnetic correlator.
	For the usual magnetic two-point function, Carrollian conformal symmetry fixes the dependence on the spatial separation to a power law \cite{Chen:2021xkw,Salzer:2023jqv,Cotler:2024xhb}.
	By contrast, in eq. \eqref{eq:TwoDimensionalGenericCorrelator}, the $z$ dependence appears inside the arbitrary function $f$ through the invariant $z^2/(\kappa_1\kappa_2)$. \par

	It is worth emphasizing that the generic non-contact expression \eqref{eq:TwoDimensionalGenericCorrelator} cannot be specialized to an operator and its conjugate.
	For $\beta_1+\beta_2=0$ and $\kappa_1+\kappa_2=0$, the expression is ill defined, and the $K^1$ Ward identity excludes a nonzero non-contact solution.
	These parameter conditions instead support the contact solution \eqref{eq:TwoDimensionalGenericContactCorrelator} described below. \par

	We next consider a correlator supported at $z=0$, which we call a contact correlator.
	The $B^1$ and $K^0$ Ward identities require $\beta_1+\beta_2=0$ and $\kappa_1+\kappa_2=0$.
	On this contact branch, the non-contact invariants no longer define independent variables.
	Instead, $u/(\kappa_1-\kappa_2)$ emerges as an invariant of the symmetry action restricted to the contact support and can therefore enter the arbitrary function $f$.
	The contact solution is
	\begin{equation}\label{eq:TwoDimensionalGenericContactCorrelator}
		\begin{aligned}
			& \mathcal{G}_{\mathrm{e}} = f\!\left( \frac{u}{\kappa_1-\kappa_2} \right) \exp\!\left( -\frac{ iu\left(\beta_1^2+\beta_2^2\right) }{ \kappa_1-\kappa_2 } \right) \abs{u}^{2-\Delta_1-\Delta_2}  \delta\!\left(z\right) \delta\!\left(\beta_1+\beta_2\right) \delta\!\left(\kappa_1+\kappa_2\right).
		\end{aligned}
	\end{equation}
	The Ward identities impose no additional selection rule on $\Delta_1$ and $\Delta_2$.
	Because electric correlators are supported at coincident spatial points \cite{Bagchi:2023fbj,Cotler:2024xhb}, we refer to \eqref{eq:TwoDimensionalGenericContactCorrelator} as the generalized electric correlator.
	The nontrivial phase and the arbitrary dependence on $u/(\kappa_1-\kappa_2)$ arise from the complete representation. \par

	For an operator $\mathcal{O}_{\Delta}$ and its conjugate $\mathcal{O}_{\Delta}^{\dagger}=\mathcal{O}_{3-\Delta}^{\prime}$, we have $\Delta_2=3-\Delta_1$, $\beta_2=-\beta_1$, and $\kappa_2=-\kappa_1$.
	The two-point function of an operator and its conjugate is therefore the contact solution
	\begin{equation}
		\begin{aligned}
			& \left\langle \mathcal{O}_{\Delta}\left(u_1,z_1;\beta,\kappa\right) \mathcal{O}_{3-\Delta}^{\prime}\left(u_2,z_2;-\beta,-\kappa\right) \right\rangle_{\mathrm{e}} = \abs{u}^{-1} f\!\left( \frac{u}{2\kappa} \right) \exp\!\left( -\frac{iu\beta^2}{\kappa} \right) \delta\!\left(z\right).
		\end{aligned}
	\end{equation}
	This result shows that Carrollian conformal symmetry permits a nonzero contact two-point function between the $\kappa>0$ and $\kappa<0$ orbits. \par

\subsubsection{Both insertions with \texorpdfstring{$\kappa_1=0$}{kappa1 zero} and \texorpdfstring{$\kappa_2=0$}{kappa2 zero}}\label{subsec:TwoDimensionalZeroKappaCorrelators}

	The second orbit configuration consists of two operators in $\kappa=0$ orbits.
	The finite transformations in Table \ref{tab:TwoDimensionalCompletedFiniteTransformations} preserve two candidate invariants
	\begin{equation}
		\mathcal{I}_1=\beta_1,\qquad \mathcal{I}_2=\beta_2.
	\end{equation} \par

	For the non-contact branch, the Ward identities fix the time dependence of the phase and impose the support $\beta_1=\beta_2$.
	The resulting non-contact two-point function is
	\begin{equation}\label{eq:TwoDimensionalZeroKappaNonContact}
		\mathcal{G}_{\mathrm{m}} = f\!\left(\beta_1+\beta_2\right) \exp\!\left( -\frac{iu\left(\beta_1+\beta_2\right)}{z} \right) \abs{z}^{-\Delta_1-\Delta_2} \delta\!\left(\beta_1-\beta_2\right) \delta_{\Delta_1,\Delta_2}.
	\end{equation}
	On this support, only $\beta_1+\beta_2$ remains as an independent argument of the free function, while the Dirac distribution implements the non-contact support.
	Here and below, the Kronecker delta $\delta_{\Delta_1,\Delta_2}$ denotes the selection rule $\Delta_1=\Delta_2$.
	The fixed power law dependence on the spatial separation identifies \eqref{eq:TwoDimensionalZeroKappaNonContact} as the magnetic non-contact correlator. \par

	For the contact branch, we start from the spatial support $z=0$.
	The remaining Ward identities then require $\beta_1=\beta_2=0$ and give
	\begin{equation}\label{eq:TwoDimensionalZeroKappaContact}
		\mathcal{G}_{\mathrm{e}} = c_{\mathrm{e}}\,\lvert u\rvert^{1-\Delta_1-\Delta_2} \delta(z)\delta(\beta_1)\delta(\beta_2).
	\end{equation}
	The contact solution \eqref{eq:TwoDimensionalZeroKappaContact} contains no arbitrary function and is fixed up to the overall coefficient $c_{\mathrm{e}}$.
	Dilatation covariance fixes the power of $\abs{u}$ without imposing an additional selection rule on $\Delta_1$ and $\Delta_2$.
	Its support at coincident spatial points identifies \eqref{eq:TwoDimensionalZeroKappaContact} as the electric contact correlator. \par

	For an operator and its conjugate in this sector, eq. \eqref{eq:TwoDimensionalZeroKappaDualOperator} gives $\beta_2=-\beta_1$ and $\Delta_2=2-\Delta_1$.
	Both the contact support and the special non-contact support described below require $\beta=0$, so no nonzero two-point function exists for a conjugate pair with generic $\beta\neq0$.
	At $\beta=\kappa=0$, the electric contact correlator is
	\begin{equation}
		\begin{aligned}
			& \left\langle \mathcal{O}_{\Delta}\left(u_1,z_1;\beta,0\right) \mathcal{O}_{2-\Delta}^{\prime}\left(u_2,z_2;-\beta,0\right) \right\rangle_{\mathrm{e}} = c_{\mathrm{e}}\,\lvert u\rvert^{-1}\delta(z)\delta(\beta).
		\end{aligned}
	\end{equation}
	The same representation parameters also admit the magnetic non-contact solution
	\begin{equation}
		\mathcal{G}_{\mathrm{m}}=c_{\mathrm{m}}\lvert z\rvert^{-\Delta_1-\Delta_2}\delta(\beta_1)\delta(\beta_2)\delta_{\Delta_1,\Delta_2}.
	\end{equation}
	The selection rule remains $\Delta_1=\Delta_2$.
	For a conjugate pair, the selection rule $\Delta_1=\Delta_2$, together with the dual relation $\Delta_2=2-\Delta_1$, fixes $\Delta_1=\Delta_2=1$.
	The spatial dependence is therefore $c_{\mathrm{m}}\lvert z\rvert^{-2}$.
	The magnetic non-contact and electric contact results at $\beta=\kappa=0$ are the 2D counterparts of the corresponding Carrollian two-point structures discussed in \cite{Chen:2021xkw,Bagchi:2023fbj,Salzer:2023jqv,Cotler:2024xhb}. \par

\subsubsection{Mixed \texorpdfstring{$\kappa$}{kappa} sectors}\label{subsec:TwoDimensionalMixedKappaCorrelators}

	The final 2D configuration is mixed.
	Without loss of generality, the second insertion lies in the $\kappa_2=0$ orbit.
	Two candidate invariants may be chosen as
	\begin{equation}\label{eq:TwoDimensionalMixedInvariants}
		\mathcal{I}_1=\beta_2,\qquad \mathcal{I}_2=\frac{z}{z\beta_1-\kappa_1}.
	\end{equation}
	\par

	For the non-contact branch, the remaining Ward identities constrain the correlator to the support
	\begin{equation}
		z = \frac{\kappa_1}{\beta_1-\beta_2}.
	\end{equation}
	On this support, the second combination in \eqref{eq:TwoDimensionalMixedInvariants} becomes $1/\beta_2$ wherever this invariant chart is regular.
	Therefore, only one independent invariant remains, which may be taken to be $\beta_2$.
	Solving the remaining Ward identities gives the non-contact correlator
	\begin{equation}\label{eq:TwoDimensionalMixedCorrelator}
		\begin{aligned}
			& \mathcal{G}_{\mathrm{m}} = f\!\left(\beta_2\right) \exp\!\left( -\frac{iu\left(\beta_1+\beta_2\right)}{z} \right) \abs{z}^{2-2\Delta_2} \abs{\kappa_1}^{-1-\Delta_1+\Delta_2} \delta\!\left( z-\frac{\kappa_1}{\beta_1-\beta_2} \right).
		\end{aligned}
	\end{equation}
	The arbitrary function therefore depends only on $\beta_2$, while the Ward identities fix the remaining phase, powers, and shifted support.
	No additional selection rule relates $\Delta_1$ to $\Delta_2$.
	The explicit power law in $z$ shows that \eqref{eq:TwoDimensionalMixedCorrelator} is the magnetic non-contact correlator for the mixed orbit configuration.
	This correlator is not invariant under the exchange of the insertion labels $1$ and $2$.
	Exchanging the insertions gives the complementary configuration with $\kappa_1=0$ and $\kappa_2\neq0$. \par

	The mixed configuration has no independent contact solution.
	Indeed, on $z=0$ the $K^0$ Ward identity requires $\kappa_1=0$, which reduces the configuration to the $\kappa_i = 0$ sector discussed in Subsection \ref{subsec:TwoDimensionalZeroKappaCorrelators}.
	Moreover, conjugation maps $\kappa$ to $-\kappa$ and therefore preserves whether $\kappa$ vanishes, so a mixed pair cannot describe an operator and its conjugate.
	The mixed orbit configuration consequently contains only the non-contact solution \eqref{eq:TwoDimensionalMixedCorrelator}. \par

\subsection{3D Carrollian Two-Point Correlation Functions}\label{subsec:ThreeDimensionalCorrelators}

	The 3D analysis follows the organization of the 2D case, using the same ordering of orbit configurations and the same sequence of steps.
	We consider in turn two operators in $\kappa\neq0$ orbits, two operators in $\kappa=0$ orbits, and the mixed configuration in which only one insertion has nonzero $\kappa$. \par

	In the contact solutions, the support conditions localize a 2D plane of coordinates or representation parameters to a point.
	We therefore also consider real blow-ups of selected point supports, with each blow-up circle retaining the angular data required by the rotation Ward identity.
	The resulting blow-up branches are additional solutions on enlarged spaces rather than alternative expressions for distributions on the original planes. 
	To specify the branches of the angular factors below, every nonzero complex factor $w$ raised to an exponent determined by spin is defined by $w^\alpha=\lvert w\rvert^\alpha e^{i\alpha\arg(w)}$, with its angle continued on the chosen $\mathrm{Spin}(2)$ lift.
	The same convention applies to the complex ratios in the coordinate and boost label planes.
	Because $2l_i\in\mathbb{Z}$, a simultaneous $4\pi$ rotation of all angular arguments leaves each complete angular factor unchanged, while its $2\pi$ monodromy cancels the central action of $\mathrm{Spin}(2)$ on the two operators.
	Global covariance therefore imposes no additional spin selection rule beyond those stated for the individual branches. \par

\subsubsection{Both insertions with \texorpdfstring{$\kappa_1\neq0$}{kappa1 nonzero} and \texorpdfstring{$\kappa_2\neq0$}{kappa2 nonzero}}\label{subsec:ThreeDimensionalGenericKappaCorrelators}

	Our analysis begins with the non-contact solution for the correlator of operators in $\kappa\neq0$ orbits.
	For two spatial vectors, we use $\vec{a}\times\vec{b}=\epsilon^{ab}a^a b^b$.
	The full space of two insertions admits four candidate invariants.
	\begin{equation}\label{eq:3DGenericInvariantCombinations}
		\begin{aligned}
			\mathcal{I}_1&=\frac{\vec{z}^{\,2}}{\kappa_1\kappa_2},\\
			\mathcal{I}_2&=\frac{\kappa_2\vec{z}\cdot\vec{\beta}_1-\kappa_1\vec{z}\cdot\vec{\beta}_2 +2(\vec{z}\cdot\vec{\beta}_1)(\vec{z}\cdot\vec{\beta}_2) -\vec{z}^{\,2}\vec{\beta}_1\cdot\vec{\beta}_2}{\vec{z}^{\,2}},\\
			\mathcal{I}_3&=\frac{\kappa_2(\vec{z}\times\vec{\beta}_1)-\kappa_1(\vec{z}\times\vec{\beta}_2) +(\vec{z}\cdot\vec{\beta}_1)(\vec{z}\times\vec{\beta}_2) +(\vec{z}\times\vec{\beta}_1)(\vec{z}\cdot\vec{\beta}_2)}{\kappa_1\kappa_2},\\
			\mathcal{I}_4&=\frac{(\kappa_1+\kappa_2-\vec{z}\cdot(\vec{\beta}_1-\vec{\beta}_2))^2 +(\vec{z}\times(\vec{\beta}_1+\vec{\beta}_2))^2}{\kappa_1\kappa_2}.
		\end{aligned}
	\end{equation}
	These four expressions are independent invariants under the Carrollian conformal transformations listed in Table \ref{tab:ThreeDimensionalCompletedFiniteTransformations}.
	\par

	The Ward identities generated by $B^a$ and $K^0$ fix the dependence of the phase on time
	\begin{equation}
		\exp\left(-\frac{iu(\vec{\beta}_1^{\,2}-\vec{\beta}_2^{\,2})}{\kappa_1+\kappa_2}\right),
	\end{equation}
	and impose two spatial constraints
	\begin{equation}
		\delta^{(2)}\!\left(\vec{z}-\frac{(\kappa_1+\kappa_2)(\vec{\beta}_1+\vec{\beta}_2)}{\vec{\beta}_1^{\,2}-\vec{\beta}_2^{\,2}}\right).
	\end{equation}
	On this support, only two independent combinations from \eqref{eq:3DGenericInvariantCombinations} remain, which may be chosen as
	\begin{equation}
		\frac{\kappa_1\vec{\beta}_2^{\,2}+\kappa_2\vec{\beta}_1^{\,2}}{\kappa_1+\kappa_2}, \qquad \frac{\vec{z}^{\,2}}{\kappa_1\kappa_2}.
	\end{equation}
	The arbitrary function of these two combinations contains the dynamical information not fixed by symmetry. \par

	For generic values of the representation parameters, solving the remaining Ward identities generated by $D$, $J^{12}$, and $K^a$ gives
	\begin{equation}\label{eq:ThreeDimensionalGenericCorrelator}
		\begin{aligned}
			& \mathcal{G}_{\mathrm{m}} = f\!\left( \frac{ \kappa_1\vec{\beta}_2^{\,2} +\kappa_2\vec{\beta}_1^{\,2} }{ \kappa_1+\kappa_2 }, \frac{\vec{z}^{\,2}}{\kappa_1\kappa_2} \right) \exp\!\left( -\frac{iu\left(\vec{\beta}_1^{\,2}-\vec{\beta}_2^{\,2}\right)} {\kappa_1+\kappa_2} \right) \abs{\kappa_1}^{1-\Delta_1} \abs{\kappa_2}^{1-\Delta_2} \\
			&\quad\times \left( \frac{ \kappa_1+\kappa_2 }{ \kappa_1 \left(\beta_2^1-i\beta_2^2\right) \left(z^1+iz^2\right) +\kappa_2 \left(\beta_1^1+i\beta_1^2\right) \left(z^1-iz^2\right) } \right)^{l_1-l_2} \frac{ \left(z^1-iz^2\right)^{l_1} }{ \left(z^1+iz^2\right)^{l_2} } \\
			&\quad\times \frac{1}{\lvert\vec{\beta}_1^{\,2}-\vec{\beta}_2^{\,2}\rvert}\delta^{(2)}\!\left(\vec{z}-\frac{(\kappa_1+\kappa_2)(\vec{\beta}_1+\vec{\beta}_2)}{\vec{\beta}_1^{\,2}-\vec{\beta}_2^{\,2}}\right).
		\end{aligned}
	\end{equation}
	The general non-contact solution imposes no additional selection rule relating $\Delta_1$ to $\Delta_2$ or $l_1$ to $l_2$.
	Like its 2D counterpart \eqref{eq:TwoDimensionalGenericCorrelator}, \eqref{eq:ThreeDimensionalGenericCorrelator} is a generalized magnetic correlator.
	Its radial dependence on $\vec{z}$ enters the arbitrary function $f$ through the invariant $\vec{z}^{\,2}/(\kappa_1\kappa_2)$ rather than being fixed to a single power law. \par

	At $\vec{\beta}_1+\vec{\beta}_2=0$ and $\kappa_1+\kappa_2=0$, the generic expression \eqref{eq:ThreeDimensionalGenericCorrelator} is not well defined.
	The $K^a$ Ward identities further exclude any non-contact solution under these parameter conditions.
	No choice of operator normalization produces a non-contact solution.
	Any nonzero limit consistent with the Ward identities would instead belong to the contact branch. \par

	The corresponding contact branch has support at $\vec{z}=0$.
	The $B^a$ and $K^0$ Ward identities then require $\vec{\beta}_1+\vec{\beta}_2=0$ and $\kappa_1+\kappa_2=0$, respectively.
	On this support, the invariant combinations in \eqref{eq:3DGenericInvariantCombinations} no longer provide independent coordinates.
	The finite Carrollian transformations listed in Table \ref{tab:ThreeDimensionalCompletedFiniteTransformations} nevertheless preserve the combination
	\begin{equation}
		\frac{2u}{\kappa_1-\kappa_2},
	\end{equation}
	which can therefore enter the arbitrary function $f$.
	In the ordinary configuration space, $\lvert\vec{z}\rvert=0$ and $\lvert\vec{\beta}_1+\vec{\beta}_2\rvert=0$ each define a single point in the corresponding 2D plane.
	No angular data associated with approaching these contact points are retained, so the orbital part of the rotation Ward identity cannot compensate the spin $l_1+l_2$.
	The contact solution on the ordinary configuration space consequently requires $l_1+l_2=0$ and takes the form
	\begin{equation}\label{eq:3DPointContact2Pt}
		\begin{aligned}
			& \mathcal{G}_{\mathrm{e}} = f\!\left( \frac{2u}{\kappa_1-\kappa_2} \right) \exp\!\left( -\frac{ iu \left(\vec{\beta}_1-\vec{\beta}_2\right)^2 }{ 2\left(\kappa_1-\kappa_2\right) } \right) \abs{u}^{3-\Delta_1-\Delta_2} \delta_{l_1+l_2,0} \delta\!\left(\kappa_1+\kappa_2\right) \delta^{(2)}\!\left(\vec{z}\right) \delta^{(2)}\!\left(\vec{\beta}_1+\vec{\beta}_2\right).
		\end{aligned}
	\end{equation}
	The power $3-\Delta_1-\Delta_2$ includes the scaling degrees of $\delta(\kappa_1+\kappa_2)$ and $\delta^{(2)}(\vec{z})$.
	The Ward identities impose no additional selection rule on $\Delta_1$ and $\Delta_2$.
	The result supported at a point in eq. \eqref{eq:3DPointContact2Pt} is the 3D counterpart of the generalized electric correlator in eq. \eqref{eq:TwoDimensionalGenericContactCorrelator}. \par

	To retain angular data at contact, we blow up the origin of the 2D $\vec{z}$ plane.
	We write
	\begin{equation}\label{eq:RealBlowUpDefinition}
		\vec{z}=r_z(\cos\phi_z,\sin\phi_z),\qquad
		\phi_z=\arg(z^1+iz^2),\qquad
		\mathbb{R}^2\to[0,\infty)_{r_z}\times S^1_{\phi_z}.
	\end{equation}
	The locus $r_z=0$ is the circle at $\lvert\vec{z}\rvert=0$, and $\delta(r_z)$ is the radial delta distribution supported on this circle.
	The angular function satisfies
	\begin{equation}
		\frac{z^1+iz^2}{z^1-iz^2}=\exp(2i\phi_z).
	\end{equation}
	The boost label support remains the ordinary vector distribution $\delta^{(2)}(\vec{\beta}_1+\vec{\beta}_2)$.
	The contact solution on this enlarged space is
	\begin{equation}\label{eq:3DBlownUpContact2Pt}
		\begin{aligned}
			& \mathcal{G}_{\mathrm{e}} = f\!\left( \frac{2u}{\kappa_1-\kappa_2} \right) \exp\!\left( -\frac{ iu \left(\vec{\beta}_1-\vec{\beta}_2\right)^2 }{ 2\left(\kappa_1-\kappa_2\right) } \right) \abs{u}^{2-\Delta_1-\Delta_2} \left( \frac{z^1+iz^2} {z^1-iz^2} \right)^{-\frac{1}{2}\left(l_1+l_2\right)} \\
			&\quad\times \delta(\kappa_1+\kappa_2) \delta(r_z) \delta^{(2)}(\vec{\beta}_1+\vec{\beta}_2).
		\end{aligned}
	\end{equation}
	This expression is a radial distribution on $[0,\infty)_{r_z}\times S^1_{\phi_z}$, not an angular function multiplied by $\delta^{(2)}(\vec{z})$ on the original plane.
	The power $2-\Delta_1-\Delta_2$ includes the scaling degrees of $\delta(\kappa_1+\kappa_2)$ and $\delta(r_z)$.
	The angular factor in \eqref{eq:3DBlownUpContact2Pt} carries the orbital rotation weight needed to cancel $l_1+l_2$, so the local Ward identities no longer require $l_1+l_2=0$.
	As in the solution on the ordinary configuration space, there is no additional selection rule on $\Delta_1$ and $\Delta_2$.
	It follows from the same Ward identity support as eq. \eqref{eq:3DPointContact2Pt}, but it is a different distribution because only the $\vec{z}$ point support has been replaced by a radial distribution carrying angular data. \par

	Using $\mathcal{O}_{\Delta,l}^{\dagger}=\mathcal{O}_{4-\Delta,-l}^{\prime}$, the relation for the dual representation in \eqref{eq:ThreeDimensionalDualOperator} imposes $\Delta_2=4-\Delta_1$, $\vec{\beta}_2=-\vec{\beta}_1$, $\kappa_2=-\kappa_1$, and $l_2=-l_1$ for an operator and its conjugate.
	Since no nonzero non-contact solution exists under these conditions, the Ward identities admit a contact correlator on each of the two configuration spaces.
	The contact correlator on the ordinary configuration space is
	\begin{equation}
		\begin{aligned}
			& \left\langle \mathcal{O}_{\Delta,l} \left( u_1,z_1^a;\beta^a,\kappa \right) \mathcal{O}_{4-\Delta,-l}^{\prime} \left( u_2,z_2^a;-\beta^a,-\kappa \right) \right\rangle_{\mathrm{e}} = \abs{u}^{-1}f\left(\frac{u}{\kappa}\right) \exp\left(-\frac{iu\vec{\beta}^{\,2}}{\kappa}\right) \delta^{(2)}(\vec{z}),
		\end{aligned}
	\end{equation}
	The corresponding contact correlator on the blow-up configuration space is
	\begin{equation}
		\begin{aligned}
			& \left\langle \mathcal{O}_{\Delta,l} \left( u_1,z_1^a;\beta^a,\kappa \right) \mathcal{O}_{4-\Delta,-l}^{\prime} \left( u_2,z_2^a;-\beta^a,-\kappa \right) \right\rangle_{\mathrm{e,bl}} = \abs{u}^{-2}f\left(\frac{u}{\kappa}\right) \exp\left(-\frac{iu\vec{\beta}^{\,2}}{\kappa}\right) \delta(r_z).
		\end{aligned}
	\end{equation}
	The angular factor becomes unity for the conjugate pair because $l_1+l_2=0$.
	The powers $\abs{u}^{-1}$ and $\abs{u}^{-2}$ follow respectively from the scaling degrees of $\delta^{(2)}(\vec{z})$ on the ordinary configuration space and $\delta(r_z)$ on the enlarged space.
	Thus the two correlators arise from the same contact conditions imposed by the Ward identities but are not two expressions for the same distribution.
	The blow-up correlator is an additional solution on the enlarged configuration space. \par

\subsubsection{Both insertions with \texorpdfstring{$\kappa_1=0$}{kappa1 zero} and \texorpdfstring{$\kappa_2=0$}{kappa2 zero}}

	We now turn to a pair of operators belonging to $\kappa=0$ orbits.
	The finite Carrollian transformations in Table \ref{tab:ThreeDimensionalCompletedFiniteTransformations} preserve two candidate invariants independently.
	\begin{equation}
		\mathcal{I}_1=\vec{\beta}_1^{\,2},\qquad \mathcal{I}_2=\vec{\beta}_2^{\,2}.
	\end{equation}
	Both combinations may therefore enter the arbitrary function. \par

	The non-contact Ward identities localize the correlator on
	\begin{equation}
		\delta(\vec{\beta}_1^{\,2}-\vec{\beta}_2^{\,2}) \delta\left(\vec{z}\times(\vec{\beta}_1+\vec{\beta}_2)\right).
	\end{equation}
	On this support, only one independent radial argument remains, and the arbitrary function can be written as $f(\lvert\vec{\beta}_1\rvert)$.
	Solving the remaining Ward identities gives the non-contact correlator
	\begin{equation}\label{eq:3DZeroKappaNonContact}
		\begin{aligned}
			& \mathcal{G}_{\mathrm{m}} = f(\lvert\vec{\beta}_1\rvert) \exp\!\left( -iu\frac{\vec{z}\cdot(\vec{\beta}_1+\vec{\beta}_2)}{\vec{z}^{\,2}} \right) \\
			&\quad\times \lvert\vec{z}\rvert^{1-\Delta_1-\Delta_2} \left( \frac{z^1+iz^2} {z^1-iz^2} \right)^{-\frac{1}{2}\left(l_1+l_2\right)} \left( \frac{\beta_1^1+i\beta_1^2} {\beta_1^1-i\beta_1^2} \right)^{-\frac{1}{4}\left(l_1-l_2\right)} \left( \frac{\beta_2^1+i\beta_2^2} {\beta_2^1-i\beta_2^2} \right)^{\frac{1}{4}\left(l_1-l_2\right)} \\
			&\quad\times \lvert\vec{\beta}_1+\vec{\beta}_2\rvert \delta(\vec{\beta}_1^{\,2}-\vec{\beta}_2^{\,2}) \delta\left(\vec{z}\times(\vec{\beta}_1+\vec{\beta}_2)\right) \delta_{\Delta_1,\Delta_2}.
		\end{aligned}
	\end{equation}
	On this support, the phase is regular throughout the non-contact region $\vec{z}\neq0$ and is manifestly rotation covariant.
	The Kronecker delta explicitly implements the selection rule $\Delta_1=\Delta_2$, while no additional relation is imposed between $l_1$ and $l_2$.
	The fixed radial power law identifies \eqref{eq:3DZeroKappaNonContact} as a magnetic non-contact correlator, while $f$ contains the information not fixed by symmetry.
	At $\kappa_1=\kappa_2=0$, the result on the ordinary configuration space in \eqref{eq:3DZeroKappaNonContact} has the magnetic power law structure familiar from 3D Carrollian CFT two-point functions \cite{Chen:2021xkw,Salzer:2023jqv,Cotler:2024xhb}, while the complete representation makes the dependence on $\vec{\beta}_i$ and its distributional support explicit. \par

	For the contact branch on the ordinary configuration space, we start from the spatial support $\vec{z}=0$.
	The remaining Ward identities impose $\vec{\beta}_1-\vec{\beta}_2=0$ and $\vec{\beta}_1+\vec{\beta}_2=0$.
	Because the ordinary planes retain no angular data at these points, the rotation Ward identity further requires $l_1+l_2=0$.
	The resulting contact solution is
	\begin{equation}\label{eq:3DZeroKappaPointContact}
		\mathcal{G}_{\mathrm{e,ord}}=c_{\mathrm{e}}\lvert u\rvert^{2-\Delta_1-\Delta_2}\delta_{l_1+l_2,0} \delta^{(2)}(\vec{z})\delta^{(2)}(\vec{\beta}_1-\vec{\beta}_2)\delta^{(2)}(\vec{\beta}_1+\vec{\beta}_2).
	\end{equation}
	The power $\abs{u}^{2-\Delta_1-\Delta_2}$ includes the scaling degree of $\delta^{(2)}(\vec{z})$, and the Ward identities impose no additional selection rule on $\Delta_1$ and $\Delta_2$. \par

	We next blow up the origins of the $\vec{z}$, $\vec{\beta}_1-\vec{\beta}_2$, and $\vec{\beta}_1+\vec{\beta}_2$ planes.
	In addition to eq. \eqref{eq:RealBlowUpDefinition}, we define
	\begin{equation}
		r_{\pm}=\lvert\vec{\beta}_1\pm\vec{\beta}_2\rvert,\qquad
		\phi_{\pm}=\arg\!\left((\beta_1^1\pm\beta_2^1)+i(\beta_1^2\pm\beta_2^2)\right).
	\end{equation}
	The contact solution on the enlarged configuration space is
	\begin{equation}\label{eq:3DZeroKappaContact}
		\begin{aligned}
			& \mathcal{G}_{\mathrm{e}} = f(\phi_- -\phi_z,\phi_+ -\phi_z)\lvert u\rvert^{1-\Delta_1-\Delta_2} \left( \frac{z^1+iz^2} {z^1-iz^2} \right)^{-\frac{1}{2}\left(l_1+l_2\right)} \delta(r_z)\delta(r_-)\delta(r_+).
		\end{aligned}
	\end{equation}
	The two angle differences $\phi_- -\phi_z$ and $\phi_+ -\phi_z$ are Carrollian invariants that emerge on this contact support.
	The power $\abs{u}^{1-\Delta_1-\Delta_2}$ includes the scaling degree of $\delta(r_z)$, while the angular factor compensates the spin $l_1+l_2$.
	The Ward identities impose no additional selection rule on $\Delta_i$ or $l_i$. \par

	For an operator and its conjugate in the $\kappa=0$ sector, eq. \eqref{eq:ThreeDimensionalZeroKappaDualOperator} gives $\Delta_2=3-\Delta_1$, $\vec{\beta}_2=-\vec{\beta}_1$, and $l_2=-l_1$.
	The support conditions in eq. \eqref{eq:3DZeroKappaPointContact} and eq. \eqref{eq:3DZeroKappaContact} then require $\vec{\beta}=0$.
	On the ordinary configuration space, the electric contact correlator becomes
	\begin{equation}
		\left\langle\mathcal{O}_{\Delta,l}\left(u_1,z_1^a;\beta^a,0\right) \mathcal{O}_{3-\Delta,-l}^{\prime}\left(u_2,z_2^a;-\beta^a,0\right)\right\rangle_{\mathrm{e,ord}} =c_{\mathrm{e}}\lvert u\rvert^{-1}\delta^{(2)}(\vec{z})\delta^{(2)}(\vec{\beta}).
	\end{equation}
	The corresponding electric contact correlator on the blow-up configuration space is
	\begin{equation}
		\begin{aligned}
			& \left\langle \mathcal{O}_{\Delta,l}\left(u_1,z_1^a;\beta^a,0\right) \mathcal{O}_{3-\Delta,-l}^{\prime}\left(u_2,z_2^a;-\beta^a,0\right) \right\rangle_{\mathrm{e,bl}} = f(\phi_- -\phi_z,\phi_+ -\phi_z)\lvert u\rvert^{-2}\delta(r_z)\delta(\lvert\vec{\beta}\rvert).
		\end{aligned}
	\end{equation}
	The different powers of $\abs{u}$ follow from the scaling degrees of the distributions supported at a point and the radial distributions, so the two electric correlators are not two expressions for the same distribution. \par

	In addition, $\vec{\beta}=\kappa=0$ admits a magnetic non-contact solution on the blow-up space of representation parameters.
	The solution below is independent of angles on this blow-up support, so the rotation Ward identity imposes $l_1=l_2$.
	\begin{equation}
		\begin{aligned}
			& \mathcal{G}_{\mathrm{m}}=c_{\mathrm{m}}\lvert\vec{z}\rvert^{-\Delta_1-\Delta_2} \left(\frac{z^1+iz^2}{z^1-iz^2}\right)^{-\frac{1}{2}\left(l_1+l_2\right)}\delta(r_-)\delta(r_+)\delta_{\Delta_1,\Delta_2}\delta_{l_1,l_2}.
		\end{aligned}
	\end{equation}
	This solution contains no arbitrary function and is fixed up to the overall coefficient $c_{\mathrm{m}}$.
	The selection rule remains $\Delta_1=\Delta_2$.
	For a conjugate pair, the selection rules $\Delta_1=\Delta_2$ and $l_1=l_2$, together with the dual relations $\Delta_2=3-\Delta_1$ and $l_2=-l_1$, fix $\Delta_1=\Delta_2=\frac{3}{2}$ and $l_1=l_2=0$.
	The radial dependence is therefore $c_{\mathrm{m}}\lvert\vec{z}\rvert^{-3}$.
	This is the 3D analogue of the $\beta=\kappa=0$ magnetic result in Subsection \ref{subsec:TwoDimensionalZeroKappaCorrelators}, and it reproduces the standard magnetic power law discussed in \cite{Chen:2021xkw,Salzer:2023jqv,Cotler:2024xhb}. \par

\subsubsection{Mixed \texorpdfstring{$\kappa$}{kappa} sectors}

	It remains to analyze the mixed pair with the first operator in a $\kappa\neq0$ orbit and the second in a $\kappa=0$ orbit.
	The configuration with $\kappa_1=0$ and $\kappa_2\neq0$ follows by exchanging the two insertions. 
	% To keep the candidate invariants compact, define
	% \begin{equation}
	% 	D_1=\kappa_1^2-2\kappa_1\vec{\beta}_1\cdot\vec{z}+\vec{\beta}_1^{\,2}\vec{z}^{\,2}.
	% \end{equation}
	The three candidate invariants may then be chosen as
	\begin{equation}
		\begin{aligned}
			\mathcal{I}_1&=\frac{\vec{z}^{\,2}}{\kappa_1^2-2\kappa_1\vec{\beta}_1\cdot\vec{z}+\vec{\beta}_1^{\,2}\vec{z}^{\,2}},\qquad \mathcal{I}_2=\frac{-\kappa_1\vec{z}\cdot\vec{\beta}_2 +2(\vec{z}\cdot\vec{\beta}_1)(\vec{z}\cdot\vec{\beta}_2) -\vec{z}^{\,2}\vec{\beta}_1\cdot\vec{\beta}_2}{\kappa_1^2-2\kappa_1\vec{\beta}_1\cdot\vec{z}+\vec{\beta}_1^{\,2}\vec{z}^{\,2}}, \qquad
			\mathcal{I}_3&=\vec{\beta}_2^{\,2}.
		\end{aligned}
	\end{equation} \par

	For generic representation parameters, the non-contact Ward identities localize the solution on
	\begin{equation}
		z^a=\frac{\kappa_1(\beta_1^a+\beta_2^a)}{\vec{\beta}_1^{\,2}-\vec{\beta}_2^{\,2}}.
	\end{equation}
	On this support, the arbitrary function $f$ may depend only on $\lvert\vec{\beta}_2\rvert$.
	Solving the remaining Ward identities fixes the phase and the remaining kinematic factors, yielding the non-contact solution for the mixed orbits.
	\begin{equation}\label{eq:ThreeDimensionalMixedCorrelator}
		\begin{aligned}
			& \mathcal{G}_{\mathrm{m}} = f(\lvert\vec{\beta}_2\rvert) \exp\!\left( -\frac{iu\left(\vec{\beta}_1^{\,2}-\vec{\beta}_2^{\,2}\right)} {\kappa_1} \right) \abs{\kappa_1}^{-\Delta_1+\Delta_2} \lvert\vec{z}\rvert^{2-2\Delta_2} \left( \frac{z^1+iz^2}{z^1-iz^2} \right)^{-l_1} \left( \frac{\beta_2^1+i\beta_2^2} {\beta_2^1-i\beta_2^2} \right)^{\frac{1}{2}\left(l_1-l_2\right)} \\
			&\quad\times \frac{1}{\lvert\vec{\beta}_1^{\,2}-\vec{\beta}_2^{\,2}\rvert}\delta^{(2)}\!\left(\vec{z}-\frac{\kappa_1(\vec{\beta}_1+\vec{\beta}_2)}{\vec{\beta}_1^{\,2}-\vec{\beta}_2^{\,2}}\right).
		\end{aligned}
	\end{equation}
	% The arbitrary function $f$ contains the information not fixed by symmetry through its dependence on $\lvert\vec{\beta}_2\rvert$.
	The solution imposes no additional selection rule relating $\Delta_1$ to $\Delta_2$ or $l_1$ to $l_2$.
	The spatial power law in \eqref{eq:ThreeDimensionalMixedCorrelator} places it in the magnetic non-contact branch for the mixed orbit configuration.
	This correlator is not invariant under the exchange of the insertion labels $1$ and $2$.
	Exchanging the insertions gives the complementary configuration with $\kappa_1=0$ and $\kappa_2\neq0$. \par

	A separate contact ansatz does not produce a correlator between mixed orbits.
	On support at $\vec{z}=0$, the $K^0$ Ward identity requires $\kappa_1+\kappa_2=0$.
	Because $\kappa_2=0$ in the mixed configuration, this condition forces $\kappa_1=0$ and reduces the problem to the sector in which both operators belong to $\kappa=0$ orbits.
	Moreover, conjugation maps $\kappa$ to $-\kappa$ and therefore preserves whether $\kappa$ vanishes, so a mixed pair cannot describe an operator and its conjugate.
	The mixed orbit configuration consequently contains only the non-contact solution \eqref{eq:ThreeDimensionalMixedCorrelator}. \par

\section{Discussion}\label{sec:Discussion}

	We constructed the complete local Carrollian conformal representation by supplementing the translation descendants with an independent chain that $K^0$ lowers to the primary.
	Temporal translation $P^0$ generates descendants annihilated by $K^0$, while the operators in eq. \eqref{eq:K0DescendantChain} form an independent chain that can be lowered by $K^0$.
	The complete family contains the boost multiplet at every level of this chain and all translation descendants with $m,n\geq0$, as summarized in eq. \eqref{eq:CompleteDescendantFamily}.
	Resumming the boost descendants and the $K^0$ chain gives generating operators labeled by $(\beta,\kappa)$ in 2D and $(\beta^a,\kappa)$ in 3D.
	Their corresponding local operators are then defined with transformation given by eq. \eqref{eq:TwoDimensionalRepresentation} and \eqref{eq:ThreeDimensionalRepresentation}.
	These labels encode the boost multiplets and $K^0$ descendant in the complete representation. \par

	% We also discussed the conjugation of a local operator. With the anti-Hermitian convention, a pairing that includes $d\kappa$ gives dual dimensions $3-\Delta$ in 2D and $4-\Delta$ in 3D.
	% On the invariant $\kappa=0$ sectors, the pairing excludes $d\kappa$ and gives $2-\Delta$ in 2D and $3-\Delta$ in 3D.
	% Hermitian conjugation reverses $\beta$ and $\kappa$ in 2D and reverses $\beta^a$, $\kappa$, and $l$ in 3D.
	% The finite transformations preserve the sign of $\kappa$, while the orbits at $\kappa=0$ are labeled by $\beta$ in 2D and by $\vec{\beta}^{\,2}$ in 3D.
	% After Fourier transformation in $u$, the quadratic Casimir takes the common form $\mathcal{C}_2=\rho\kappa-\beta^2$ or $\mathcal{C}_2=\rho\kappa-\vec{\beta}^{\,2}$, as shown in eq. \eqref{eq:2DQuadraticCasimir} and \eqref{eq:3DQuadraticCasimir}.
	% For $\kappa\neq0$, the higher Casimir in eq. \eqref{eq:3DQuarticCasimir} and its 2D counterpart act by derivatives, so these orbit representations are reducible.
	% At $\kappa=0$, the Casimirs are constant on each fixed orbit.
	% The stabilizer analysis gives an irreducible orbit representation when no additional internal labels are present.
	% The known 2D representations with boost charge are recovered at $\kappa=0$, while the standard 3D representation used in 4D flat holography lies at $\vec{\beta}=\kappa=0$. \par

	We also determined the dual representations, orbit decomposition, and Casimir structure.
	When $\kappa$ is an independent label, the pairing includes $d\kappa$ and gives dual dimensions $3-\Delta$ in 2D and $4-\Delta$ in 3D, whereas the restricted pairing on the invariant $\kappa=0$ sectors excludes $d\kappa$ and gives $2-\Delta$ and $3-\Delta$, respectively.
	Hermitian conjugation reverses $\beta$ and $\kappa$ in 2D and reverses $\beta^a$, $\kappa$, and $l$ in 3D.
    Therefore the conjugate operators are given by
    \begin{equation}
        (\mathcal{O}_{\Delta}^{\mathrm{(2D)}}(u,z;\beta,\kappa))^\dagger=\mathcal{O}_{3-\Delta}^{\prime~\mathrm{(2D)}}(u,z;-\beta,-\kappa), \quad
        (\mathcal{O}_{\Delta, l}^{\mathrm{3D}}(u,z^a;\beta^a,\kappa))^\dagger=\mathcal{O}_{4-\Delta,-l}^{\prime~\mathrm{(3D)}}(u,z^a;-\beta^a,-\kappa),
    \end{equation}
    respectively.
	Since finite transformations preserve the sign of $\kappa$, the regions with $\kappa>0$ and $\kappa<0$ form two distinct orbits exchanged by conjugation.
	At $\kappa=0$, the orbits are labeled by a fixed $\beta$ in 2D and a fixed $\vec{\beta}^{\,2}$ in 3D.
	After Fourier transformation in $u$, the quadratic Casimir takes the common form $\mathcal{C}_2=\rho\kappa-\beta^2$ in 2D and $\mathcal{C}_2=\rho\kappa-\vec{\beta}^{\,2}$ in 3D, as shown in eq. \eqref{eq:2DQuadraticCasimir} and eq. \eqref{eq:3DQuadraticCasimir}.
	For $\kappa\neq0$, the Pauli--Lubanski Casimir in 2D and the 3D quartic Casimir in eq. \eqref{eq:3DQuarticCasimir} contain derivative terms, so the corresponding orbit representations are reducible.
	At $\kappa=0$, the Casimirs are constant on each fixed orbit, and the stabilizer analysis gives an irreducible orbit representation when no additional internal labels are present.
	The known 2D representations with boost charge are recovered on the $\kappa=0$ orbits, while the standard 3D representation used in 4D flat holography lies at $\vec{\beta}=\kappa=0$. \par

	We solved the global two-point Ward identities for the three possible orbit configurations.
	In every case, the Ward identities fix the support, phases, homogeneous factors, and selection rules, while any remaining function depends only on Carrollian invariants.
	When both insertions have nonzero $\kappa$, eq. \eqref{eq:TwoDimensionalGenericCorrelator} and \eqref{eq:ThreeDimensionalGenericCorrelator} give non-contact branches on shifted support, while eq. \eqref{eq:TwoDimensionalGenericContactCorrelator} and \eqref{eq:3DPointContact2Pt} give independent contact branches.
	The contact branches impose no additional relation between $\Delta_1$ and $\Delta_2$.
	When both insertions have $\kappa=0$, the non-contact branches have the fixed spatial power laws in eq. \eqref{eq:TwoDimensionalZeroKappaNonContact} and \eqref{eq:3DZeroKappaNonContact}.
	A conjugate pair with nonzero boost labels has a vanishing two-point function, whereas the sector with zero boost labels admits both an electric contact branch and an additional magnetic non-contact branch.
	For a conjugate magnetic pair, these solutions require $\Delta=1$ in 2D and $\Delta=\frac{3}{2}$ in 3D.
	The 3D solution also requires $l=0$.
	A mixed pair with only one nonzero $\kappa$ admits only the magnetic non-contact solutions in eq. \eqref{eq:TwoDimensionalMixedCorrelator} and \eqref{eq:ThreeDimensionalMixedCorrelator}. \par

	In 3D, ordinary point support retains no angular data.
	For the nonzero $\kappa$ contact branch, the real blow-up replaces only the $\vec{z}$ point support by a circle, as in eq. \eqref{eq:3DBlownUpContact2Pt}.
	For the zero $\kappa$ contact branch, the three circles in eq. \eqref{eq:3DZeroKappaContact} admit an arbitrary function of two invariant angle differences.
	These solutions are distinct distributions on enlarged blow-up spaces rather than alternative expressions for distributions supported at points on the original planes.
	The $\vec{\beta}=\kappa=0$ sector also defines an invariant submodule whose invariants and support differ from those of the generic orbits.
	Its two-point correlation functions must therefore be solved directly and cannot in general be obtained by setting the labels to zero in a generic expression.
	Our analysis is limited to global symmetry, the specified spin content in 2D and 3D, and two-point correlation functions.
	The algebraic Hermiticity conditions do not by themselves establish a positive definite physical Hilbert space.
	A further decomposition of the reducible sectors with $\kappa\neq0$ and explicit realizations of the complete representation require additional physical input. \par

	The results above leave several open questions that motivate the directions discussed below.
	A central problem is to construct explicit Carrollian field theories that realize the complete representation, particularly its sectors with $\kappa\neq0$.
	Such theories could determine which two-point structures occur dynamically and constrain the arbitrary functions left undetermined by the global Ward identities.
	Higher-point correlation functions and crossing relations provide additional consistency conditions.
	Operator product expansions \cite{Pate:2019lpp,Nguyen:2025sqk} and Ward identities in momentum space \cite{Marotta:2025qjh} offer systematic tools for deriving these constraints. 
	Differential representations provide a practical starting point for these calculations \cite{Chakrabortty:2024bvm}.
	The generator actions derived here allow the analysis to be organized separately for each orbit configuration and make it possible to study how products of operators from different sectors decompose.
	Carrollian limits of AdS correlators and scattering equation methods offer complementary approaches to contact and exchange contributions \cite{Surubaru:2025fmg,Adamo:2025bfr}.
	Applying these methods to the $\kappa\neq0$ and mixed orbit configurations would connect the kinematic classification established here to explicit Carrollian dynamics. \par

	% Several directions follow from the present work.
	% The arbitrary functions left undetermined by the two-point Ward identities define a natural dynamical problem beyond the present analysis.
	% Higher-point correlation functions, crossing relations, and explicit Carrollian field theories can constrain this functional freedom.
	% Operator product expansions \cite{Pate:2019lpp,Nguyen:2025sqk} and Ward identities in momentum space \cite{Marotta:2025qjh} provide complementary tools.
	% Differential representations offer a practical starting point for these calculations \cite{Chakrabortty:2024bvm}.
	% Carrollian limits of AdS correlators and scattering equation methods offer alternative approaches to contact and exchange contributions \cite{Surubaru:2025fmg,Adamo:2025bfr}.
	% The generator actions derived here make it possible to formulate these questions separately for each orbit configuration and to study how products of different sectors decompose. \par

	The representation theory can also be extended beyond the global 2D and 3D algebras considered here.
	In higher dimensions, the boost descendants must be organized into representations of the full spatial rotation group, and the independent higher Casimirs and orbit stabilizers must be classified \cite{Chen:2021xkw,Afshar:2024llh,Kulkarni:2025qcx}.
	For infinite dimensional, boundary, or supersymmetric extensions, one must determine whether the additional generators preserve the $K^0$ chain or require new descendant directions \cite{Bagchi:2024qsb,Zheng:2025cuw,Buzaglo:2025nti}.
	A comparison with celestial conformal multiplets may clarify whether relations among celestial descendants have an analogue in the complete Carrollian conformal representation \cite{Pasterski:2021fjn}. \par

	% A separate question is which algebraically allowed sectors occur in explicit theories.
	% Bosonic and fermionic Carrollian field theories provide explicit settings in which this question can be tested \cite{Hao:2021urq,Hao:2022xhq}.
	% Algebraic constructions of Carrollian quantum states provide another test \cite{Fredenhagen:2026pia}.
	% Holographic limits from AdS \cite{Lipstein:2025jfj} and holographic renormalization \cite{Ammon:2025jmy} offer further tests of the source, response, and state interpretations of these sectors.
	% Such constructions must identify a positive physical inner product and the physical subspace inside the reducible sectors with $\kappa\neq0$.
	% They should also clarify the operator meaning of the contact branches and of the angular data retained by the real blow-up.
	% This analyses would determine which orbits of the complete representation occur in flat holography. \par

    A separate direction for future work is to determine which algebraically allowed sectors of the complete representation admit consistent physical realizations in explicit Carrollian theories and in flat holography.
	Bosonic and fermionic Carrollian field theories provide concrete settings in which this question can be investigated \cite{Hao:2021urq,Hao:2022xhq}, while algebraic constructions of Carrollian quantum states offer a complementary approach to the corresponding state spaces \cite{Fredenhagen:2026pia}.
	Holographic limits from AdS \cite{Lipstein:2025jfj} and holographic renormalization \cite{Ammon:2025jmy} can further test whether these sectors admit consistent interpretations in terms of sources, responses, and states.
	A physical realization must go beyond algebraic consistency by identifying a positive inner product and a physical subspace within the reducible sectors with $\kappa\neq0$.
	It should also clarify the operator meaning of the contact branches and of the angular data retained by the real blow-up.
	Together, these analyses would determine which orbits of the complete representation are physically realized and which of their correlation structures occur in flat holography. \par

	The relation $m^2=\mathcal{C}_2=\kappa\rho-\vec{\beta}^{\,2}$ in eq. \eqref{eq:MassiveCasimirRelation} shows why the sectors with $\kappa\neq0$ can support a boundary description of massive particles with real labels.
	Massive conformal bases have been constructed for scalar particles, spinning bosons, and fermions \cite{Pasterski:2016qvg,Law:2020tsg,Narayanan:2020amh}.
	Massive Carrollian fields at timelike infinity provide a complementary description \cite{Have:2024dff}.
	Induced representations provide a framework for comparing the orbit classification found here with massive Poincar\'e representations \cite{Liu:2026toc}.
	An explicit dictionary should relate Poincar\'e states to the Carrollian labels $(\beta^a,\kappa)$ and determine how these labels enter observables at null infinity.
	The same framework should connect soft graviton Ward identities to amplitudes with massive hard states \cite{Campiglia:2015kxa}.
	Some of these questions are discussed in \cite{workinprogress}.
	% The treatment of states containing several massive particles, together with factorization and loop corrections, remains open beyond that construction. \par

\section*{Acknowledgments}

	We thank Reiko Liu, Jiang Long, and Wen-Jie Ma for valuable discussions. \par

\bibliographystyle{JHEP}
\bibliography{refs}

\end{document}